\documentclass[useAMS,usenatbib,usegraphicx]{mn2e}

\usepackage{lscape}
\usepackage{times}

\def\mnras{MNRAS}
\def\aap{A\&A}
\def\aaps{A\&AS}
\def\aj{AJ}
\def\araa{ARA\&A}
\def\apjl{ApJ}
\def\apj{ApJ}
\def\apjs{ApJS}

\def\nat{Nature}
\def\pasj{PASJ}

\newcommand{\ms}{MS\,0451$-$03}
\newcommand{\ea}{et al.}
\newcommand{\HST}{{\it HST}}
\newcommand{\Spitzer}{{\it Spitzer}}
\newcommand{\myr}{${\rm M_{\sun}yr^{-1}}$}
\newcommand{\cl}{Cl\,0024+16}

\title[A 1.1-mm Survey for ULIRGs in the field of MS\,0451-03]{An AzTEC 1.1-mm Survey for ULIRGs in the field of the Galaxy Cluster MS\,0451.6$-$0305}
\author[J. L. Wardlow et al.]{
\parbox[t]{\textwidth}{%\vspace{-1cm}
J.\,L.\ Wardlow$^{1}$\thanks{E-mail: j.l.wardlow@durham.ac.uk},
Ian\ Smail$^2$, G.\,W.\ Wilson$^3$, M.\,S.\ Yun$^3$, K.\,E.\,K.\ Coppin$^2$, 
R.\ Cybulski$^3$, J.\,E.\ Geach$^1$, R.\,J.\ Ivison$^{4,5}$, I.\ Aretxaga$^6$, J.\,E.\ Austermann$^7$, A.\,C.\ Edge$^2$, 
G.\,G.\ Fazio$^8$, J.\ Huang$^8$, D.\,H.\ Hughes$^6$, T.\ Kodama$^9$, Y.\ Kang$^{10}$, S.\ Kim$^{10}$,
P.\,D.\ Mauskopf$^{11}$, T.\,A.\ Perera$^{12}$, K.\,S.\ Scott$^{3,13}$}\\\\
$^{1}$Department of Physics, Durham University, South Road, Durham, DH1 3LE, UK\\
$^{2}$Institute for Computational Cosmology, Durham University, South Road, Durham, DH1 3LE, UK\\
$^{3}$Department of Astronomy, University of Massachusetts, Amherst, MA 01003, USA\\
$^{4}$Scottish Universities Physics Alliance, Institute for Astronomy, University of Edinburgh, Blackford Hill, Edinburgh, EH9 3HJ, UK \\
$^{5}$UK Astronomy Technology Centre, Royal Observatory, Blackford Hill, Edinburgh EH9 3HJ, UK \\
$^{6}$Instituto Nacional de Astrof{\'i}sica, {\'O}ptica y Electr{\'o}nica, Tonantzintla, Puebla, M{\'e}xico\\
$^{7}$Center for Astrophysics and Space Astronomy, University of Colorado, Boulder, CO 80309, USA\\
$^{8}$Harvard-Smithsonian Center for Astrophysics, 60 Garden St. MS-65, Cambridge, MA 02138-1516, USA \\
$^{9}$National Astronomical Observatory of Japan, Mitaka, Tokyo 181-8588, Japan\\
$^{10}$Astronomy \& Space Science Department, Sejong University, Seoul, South Korea\\
$^{11}$School of Physics \& Astronomy, Cardiff University, Queens Buildings, The Parade, Cardiff, CF24 3AA, UK \\
$^{12}$ Illinois Wesleyan University, P.O. Box 2900, Bloomington, IL 61702-2900, USA \\
$^{13}$ Department of Physics and Astronomy, University of Pennsylvania, Philadelphia, PA 19104, USA}

\begin{document}

\date{Accepted year month day. Received year month day; in original form year month day}
\pagerange{\pageref{firstpage}--\pageref{lastpage}} \pubyear{2009}

\maketitle \label{firstpage}

\begin{abstract}
We have undertaken a deep ($\sigma\sim 1.1$ mJy) 1.1-mm survey of the $z=0.54$ cluster 
MS\,0451.6$-$0305 using the AzTEC camera on the James Clerk Maxwell Telescope.
We detect 36 sources with S/N$\ge 3.5$ in the central 0.10 deg$^2$ 
and present the AzTEC map, catalogue and number counts. 
We identify counterparts to 18 sources (50\%) using radio, mid-infrared, \Spitzer\ IRAC and 
Submillimeter Array data. Optical, near- and mid-infrared spectral energy 
distributions are compiled for the 14 of these galaxies with detectable counterparts,
which are expected to contain all likely cluster members. 
We then use photometric redshifts and colour selection to separate background galaxies 
from potential cluster members and test
the reliability of this technique using archival observations of submillimetre galaxies.
We find two potential MS\,0451$-$03 members, which, if they are both cluster 
galaxies have a total star-formation rate (SFR) of $\sim100$ \myr\ -- a significant fraction of 
the combined SFR of all the other galaxies in MS\,0451$-$03. 
We also examine the stacked rest-frame mid-infrared, millimetre and radio emission of cluster 
members below our AzTEC detection limit and find that the SFRs of mid-IR selected galaxies in the 
cluster and redshift-matched field populations are comparable. 
In contrast, the average SFR of the morphologically classified
late-type cluster population is $\sim 3$ times less than the corresponding redshift-matched
field galaxies. This suggests that these galaxies may be in the process of being
transformed on the red-sequence by the cluster environment.
Our survey demonstrates that although the environment of \ms\ appears to
suppress star-formation in late-type galaxies, it can support
active, dust-obscured mid-IR galaxies and potentially millimetre-detected LIRGs.
\end{abstract}


\section{Introduction}
\label{sec:intro}

Galaxy clusters are highly biased environments in which galaxies potentially 
evolve more rapidly than in the field. The galaxy populations of local massive clusters contain mainly early-type 
galaxies which define a colour-magnitude relation (CMR) \citep{Visvanathan77, Bower92}. However,
studies of clusters out to $z\sim 1$ suggest that they contain 
increasing activity at higher redshifts due to a growing fraction of blue, star-forming galaxies \citep{Butcher84}. 
Over the same redshift range there appears to be a growing deficit in the CMR population 
at faint magnitudes, as well as a claimed increasing decline in the numbers of S0 galaxies, suggesting that the blue,
star-forming galaxies may be transforming into these passive populations with time \citep{Dressler97, Smail98a, DeLucia07, Stott07, Holden09}. 

The blue, star-forming populations within the clusters are accreted from the surrounding field as the clusters grow via gravitational collapse.
The evolution in the star-forming fraction in the clusters may thus simply track the increasing activity in the field population at higher redshifts. 
The increasing activity in the field is also reflected in an increasing number of the most luminous (and dusty) star-burst galaxies with redshift
\citep[e.g.][]{LeFloch05}: the Luminous Infrared Galaxies (with L$_{\rm FIR}\geq 10^{11}$\,L$_\odot$) and their Ultraluminous cousins (ULIRGs, L$_{\rm FIR}\geq 10^{12}$\,L$_\odot$).
These systems will also be accreted into the cluster environment along with their 
less-obscured and less-active population as the clusters grow.  Indeed, mid-infrared (mid-IR) 
surveys of clusters have identified a population of dusty star-bursts whose activity increases with redshift  \citep[e.g.][]{Geach06, Geach08}.  
However, these mid-IR surveys can miss the most obscured (and potentially most active) galaxies which are optically thick in the rest-frame mid-IR.  
If they are present in clusters -- even in relatively low numbers -- such active galaxies will  
contribute significantly to the star formation rate (SFR) in these environments and the metal enrichment of the intracluster medium.
Hence to obtain a complete census of the star formation within clusters we need to survey these 
systems at even longer wavelengths, in the rest-frame far-infrared (far-IR), corresponding to the observed sub-millimetre and millimetre waveband.

Over the past decade or more there have been a number of studies of rich clusters of galaxies in the sub-millimetre and millimetre wavebands
\citep[e.g.][]{Smail97, Zemcov07, Knudsen08}. 
Many of these studies were seeking to exploit the clusters as gravitational telescopes to aid 
in the study of the distant sub-millimetre galaxy (SMG) population behind the clusters, while 
others focused on the detection of the Sunyaev-Zel'dovich (SZ) emission.
Due to the limitations of current technologies direct detection of millimetre sources is restricted to those
with fluxes S$_{\rm 1100\umu m}\ga 1$\,mJy, or equivalently galaxies with SFRs $\ga 300$\,M$_{\sun}$\,yr$^{-1}$ 
-- much higher than expected for the general cluster populations based on optical surveys.   
Nevertheless these studies have serendipitously detected a number of cluster galaxies, although these are either atypical (e.g.\ 
central cluster galaxies, \citet{Edge99}) or are not confirmed members \citep[e.g.][]{Best02, Webb05}.
More critically,  with few exceptions these studies have all focused on the central 2--3 arcmin 
of the clusters, where the SZ emission and lensing amplification both peak, and have not surveyed 
the wider environment of the cluster outskirts where much of the obscured star formation is likely to be occurring \citep[e.g.][]{Geach06}.
The two exceptions are the wide-field survey of the $z\sim0.25$ cluster A\,2125 by \citet{Wagg09} and the 
survey of an overdense region of the COSMOS field by \citet{Scott08} and \citet{Austermann09a}. 
\citet{Wagg09} detected 29 millimetre galaxies across a $\sim 25$-arcmin 
diameter region centered on A\,2125 of which none are claimed to be members. However, 
the only redshift estimates available are based on crude radio-to-millimetre 
spectral indices, which are sensitive to both dust temperature and redshift \citep{Blain03}. 
The AzTEC/COSMOS survey \citep{Scott08, Austermann09a} covered a number of structures, including
an X-ray detected $z=0.73$ cluster and concluded that the statistical overdensity of sources was most likely
due to the gravitational lensing of background SMGs by these foreground structures.

To help to conclusively answer the issue of the obscured star-forming populations in distant clusters we have therefore undertaken a
wide-field 1.1-mm survey of the $z=0.54$
rich cluster MS\,0451.6$-$0305 (hereafter \ms) with the AzTEC camera  \citep{Wilson08} on the James Clerk Maxwell Telescope (JCMT).   
This panoramic millimetre survey can also take advantage 
of the significant archival data available for this well-studied region. 
In particular the panoramic imaging of \ms\ from {\it Spitzer} and uniquely, the {\it Hubble Space Telescope} ({\it HST}), as well as 
extensive archival multiwavelength imaging and spectroscopy, which 
we employ for determining cluster membership of AzTEC-detected galaxies.
Our survey probes the rest-frame 700\,$\umu$m emission of cluster galaxies -- in 
search of examples of strongly star-forming but obscured galaxies -- as well as identifying more luminous and more distant examples of
the SMG population.   We can compare our millimetre search for cluster members to the previous mid-IR survey of this cluster 
by \citet{Geach06} who uncovered a population of dusty star-forming galaxies which dominate the integrated SFR of the
cluster of $200\pm 100$\,M$_{\sun}$\,yr$^{-1}$. 
Our AzTEC map covers $\sim 60$ times the area of the 850\,$\umu$m SCUBA 
observations of the central region of \ms\ \citep{Borys04a}, while
the depth of $\sigma\sim1.1$ mJy is sufficient to identify ULIRGs individually 
and obtain stacked constraints on far-IR fainter cluster members.

We describe our observations and the data reduction in \S\ref{sec:obsred}; 
present 1.1-mm number counts, identify 
counterparts to the AzTEC galaxies, and use photometric techniques to determine
cluster membership in \S\ref{sec:analysis}.
\S\ref{sec:conc} contains our conclusions; individual AzTEC galaxies are presented in Appendix \ref{sourcenotes}.
Throughout this paper we use J2000 coordinates and $\Lambda$CDM cosmology with $\Omega_M=0.3$, $\Omega_{\Lambda}=0.7$ and $H_0=70$ ${\rm kms^{-1}Mpc^{-1}}$
and all photometry is on the AB system unless otherwise stated.

\section{Observations and Data Reduction}
\label{sec:obsred}

\subsection{AzTEC Observations}
\label{sec:aztec}

\begin{figure*}
\begin{minipage}{17.5cm}
\includegraphics[width=17.5cm]{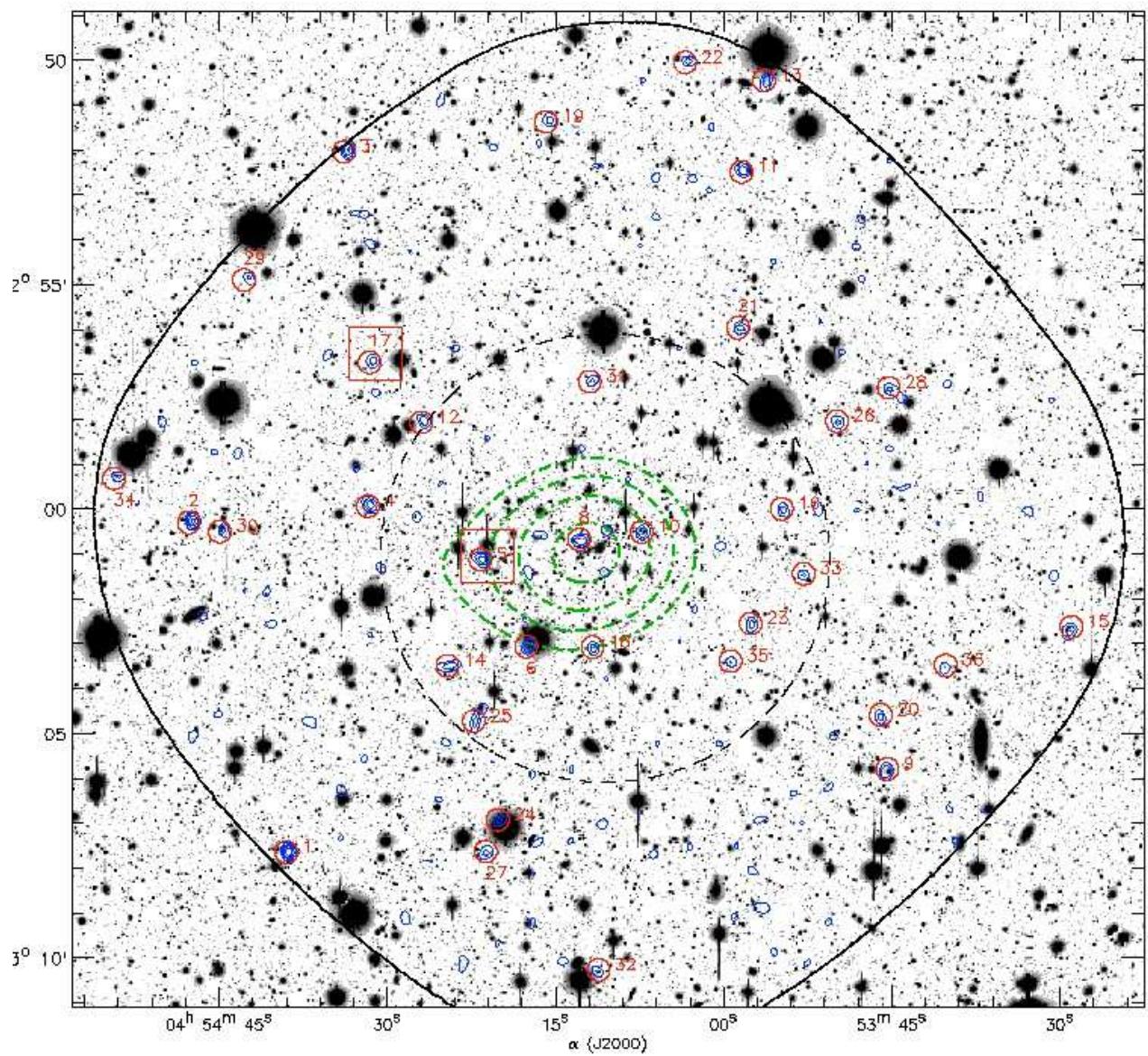}
\caption{AzTEC S/N contours for the \ms\ field shown over the corresponding Subaru {\it R}-band image; 
smoothed (dot-dashed) X-ray contours highlight the cluster centre. We also plot a 5-arcmin
radius circle (dashed) illustrating the sub-region analysed in \S\ref{sec:azsource}.
The solid black line encircles the region with exposure coverage $\ge 50\%$ of the maximum value, where sources are extracted.
In this region AzTEC contours shown are at $1\sigma$ intervals, starting at $2.5\sigma$.
15 arcsec radius circles mark the 1.1-mm source detections labelled in order of decreasing S/N (Table \ref{tab:smgs}) and squares 
highlight sources which are potential cluster members (\S\ref{sec:aztecphotoz}). 
MMJ\,045413.35 is not shown because it lies outside of the area covered by the Subaru image.
The 1.1-mm sources appear evenly distributed, with no obvious overdensity towards
the cluster centre.}
\label{fig:smgmap}
\end{minipage}
\end{figure*}

The AzTEC millimetre camera \citep{Wilson08} was installed on the JCMT
during November 2005 and operated nearly continuously
until February 2006.  On the JCMT AzTEC has a field-of-view of $\sim
18$-arcmin$^2$ and a beam of 18-arcsec FWHM.  
For this project we observed a 0.23-deg$^2$ region centred on
\ms\ at $04^{\rm h}54^{\rm m}10\fs1$,
$\-03\degr01\arcmin07\farcs6$, over a total of 24.5 hours
between 2005 December 14 and 2006 January 6. Observations were carried
out in raster-scan mode with 1200$^{\prime\prime}$ long scans in
elevation separated by 9$^{\prime\prime}$ steps in azimuth (see
\citet{Wilson08} for details of our standard raster scan strategy).
Scans were made at a fixed speed of 120$^{\prime\prime}s^{-1}$.  Data from
the turnarounds were excised for this analysis.  Each observation took
35 minutes to complete and a total of 42 observations were made of the
field.  Over the set of observations the zenith opacity at 225~GHz,
$\tau_{225}$, was monitored with the Caltech Submillimeter
Observatory's opacity meter and ranged from 0.03 to 0.17.  The
average opacity for the set of observations was 0.09, equivalent to an
atmospheric column of precipitable water vapour of $\sim2$ mm.

The 42 individual AzTEC observations were reduced and combined using
the publicly available AzTEC Data Reduction Pipeline V1.0 -- a set of
{\sc IDL} code that is optimised for detecting point sources in
blank-field AzTEC surveys.  This is the same pipeline that has been
used in previous blank-field AzTEC/JCMT analyses including
\citet{Scott08}, \citet{Perera08} and \citet{Austermann09b}.

Since the reduction of these observations follows the same procedure
and uses the same code as previous AzTEC studies, we refer the reader
to \citet{Scott08} for a thorough review of the reduction steps.  Here
we focus on the particulars of this field and the measured
characteristics of the \ms\ map.

Since we are interested in counterparts to our detected
SMGs it is critical that we understand the pointing
accuracy of the maps.  In the reduction pipeline, fine corrections to
the JCMT's pointing model are applied to each observation based on
regular observations of a nearby QSO ($\alpha=04^{\rm h}23^{\rm m}15\fs8,
\delta=-01\degr20\arcmin33\farcs1$) in the same manner as described in
\citet{Scott08} and \citet{Wilson08}.  The residual astrometric error for
the field is measured by stacking the AzTEC map at the positions of
radio sources in the field (see \S\ref{sec:multiwave}) and fitting for the centroid of the
resulting image. The stacked radio sources are detected at $\sim6\sigma$ 
and demonstrate an overall systematic astrometric shift in the AzTEC map
of $\Delta\alpha=-3.0^{\prime\prime}\pm0.9^{\prime\prime},
\Delta\delta=1.2^{\prime\prime}\pm1.5^{\prime\prime}$ with respect to
the radio reference frame. This correction is 
negligible in Dec, but the significant RA offset
is applied to the AzTEC data prior to further analysis and source extraction.

The AzTEC data are flux calibrated as described in \citet{Wilson08} from
nightly observations of Uranus over the JCMT run.  The final calibration
accuracy is $\sim10\%$.

The primary products which come out of the AzTEC pipeline are 
a map representing the average filtered point source response, a map of the 
of the weight of each pixel in the sky flux map (representing the uncertainty in flux of each pixel), and a map of 
the signal to noise estimate in each pixel.
 All maps are made on the same
$3^{\prime\prime}\times3^{\prime\prime}$ grid and all maps have been
Wiener filtered to optimize sensitivity to point sources.   
The central $\sim0.10$ deg$^2$ of the AzTEC map is relatively uniform, with
all pixels having $\ge50\%$ of the peak weight.
Sources are extracted in the central region of the maps (\S\ref{sec:azsource}) 
where the rms noise is  $\sim1.1$ mJy.
Contours of `signal-to-noise' (S/N), overlayed on the Subaru R-band 
image of the field are shown in Fig. \ref{fig:smgmap}.

\subsection{Multi-wavelength data}
\label{sec:multiwave}

\begin{table*}  
\begin{minipage}{95mm} 
\caption{Summary of the observations}
\setlength{\tabcolsep}{1mm}
\begin{tabular}{ccccl}
\hline
 Filter& Instrument& Detection & Reference \\
 && Limit \footnote{1.1 mm is the $3.5\sigma$ minimum catalogue limit in the roughly constant noise region of the map; 
in the other bands we quote $3\sigma$ limits.}\\
\hline
{\it U} & CFHT -- MegaPrime & 25.1 mag & \citet{Donovan07}\\
{\it B} & Subaru -- Suprime-Cam  & 26.6 mag & \citet{Kodama05}\\
{\it V} & Subaru -- Suprime-Cam  & 25.8 mag & \citet{Kodama05}\\
{\it R} & Subaru -- Suprime-Cam  & 25.1 mag & \citet{Kodama05}\\
{\it I} & Subaru -- Suprime-Cam  & 24.2 mag & \citet{Kodama05}\\
{$z'$} & WHT -- PFIP & 24.2 mag & This work\\
{\it K} & Palomar Hale -- WIRC &  20.1 mag & G. Smith et al., in prep. \\
3.6\,$\umu$m &{\it Spitzer} -- IRAC & 23.0 mag & This work\\
4.5\,$\umu$m &{\it Spitzer} -- IRAC & 23.2 mag & This work \\
5.8\,$\umu$m &{\it Spitzer} -- IRAC & 21.4 mag & This work\\
8.0\,$\umu$m &{\it Spitzer} -- IRAC & 22.1 mag & This work\\
24\,$\umu$m &{\it Spitzer} -- MIPS & 120 $\umu$Jy & \citet{Geach06}\\
1.1\,mm & JCMT -- AzTEC &  3.6 mJy & This work\\
1.4\,GHz & VLA & 51 $\umu$Jy & This work \\
\hline
\end{tabular}
\label{tab:allobs}
\end{minipage}
\end{table*}

Significant multiwavelength archival data exist for the \ms\ field, which 
are summarised in Table \ref{tab:allobs} and described below. 

\subsubsection{\Spitzer\ MIPS imaging}
\citet{Geach06} obtained {\it Spitzer} MIPS 24$\umu$m observations of a 
0.23-deg$^2$ area of \ms, excluding the central 25-arcmin$^2$. 
The full details of the reduction and 
source extraction process are described in \citet{Geach06}. The central 
region of the cluster was part of Guaranteed Time
Observations ({\it Spitzer} program 83) so these data were obtained from 
the archive and incorporated into the mosaic. The 5-$\sigma$ catalogue 
detection limit corresponds to 200 $\umu$Jy.

\subsubsection{Radio Imaging}
Archival observations of the \ms\ field at 1.4 GHz were obtained using the 
National Radio Astronomy 
Observatory's (NRAO\footnote{NRAO is operated by Associated 
Universities Inc., under a cooperative agreement with the National 
Science Foundation.}) Very Large Array (VLA), combining 9\,hours of data 
obtained in 2002 June in the VLA's BnA configuration with 16\,hours of 
A-configuration data taken in 2006 February (Project IDs AN109 and 
AB1199, respectively). The nearby calibrator, 0503+020, was used to 
track amplitude and phase, with absolute flux and bandpass calibration 
set via 0137+331. The now-standard 1.4-GHz wide-field imaging approach 
was adopted 
\citep{Owen05, Biggs06}, using 
spectral-line mode `4' to acquire data with an integration time of 
3\,s (10\,s in BnA) and using the {\sc imagr} task in the Astronomical 
Image Processing System ({\sc aips}) to map out the primary beam 
using a mosaic of 37 images, with 17 more distant radio sources 
covered by additional facets. Several iterations of self-calibration 
and imaging resulted in a noise level of 11\,$\mu$Jy\,beam$^{-1}$, 
with a $\rm 2.3''\times 1.8''$ synthesised beam at a position angle 
0.0$^{\circ}$.

Sources and corresponding fluxes are obtained from {\sc SExtractor} 
\citep{Bertin96}, using the S/N map for detection and the 
flux map as the analysis image; sources are extracted where a minimum 
of 10 contiguous pixels (each is 0.16 arcsec$^2$) have S/N$\ge2$.
The resulting catalogue has a 3-$\sigma$ flux limit of 51 $\umu$Jy and is 
corrected for bandwidth smearing.
We verify the statistical properties of our catalogue by comparing source
counts with \citet{Biggs06}. Notably, the source density is not significantly
enhanced towards the cluster centre, allowing us to use the counts across the 
whole field in our statistical calculation of AzTEC counterparts (\S\ref{sec:ids}).

\subsubsection{Optical and near-infrared imaging}
Observations of \ms\ in the {\it U}-band were taken with the MegaPrime 
camera on the Canada France Hawaii Telescope (CFHT) and reach a 3-$\sigma$ detection
limit of 25.1 mag in a 0.25-deg$^2$ field.
Standard reduction techniques were employed, customised for 
MegaPrime data (C.-J. Ma \& H. Ebeling, private communication), 
details of which can be found in \citet{Donovan07}.

In addition we observed a $0.08$-deg$^2$ area centered on the cluster core
on 2007 October 09 through the $z'$ filter
with the Prime Focus Imaging Platform mounted on the 4.2-m William 
Herschel Telescope (WHT). A total integration time of one hour was obtained 
and reduced using 
standard techniques and calibrated using a Sloan Digital Sky Survey 
standard field, yielding a 3-$\sigma$ limiting magnitude of 24.2 mag. 

Finally, the {\it BVRI} imaging employed here comes from the
PISCES survey \citep[Panoramic Imaging and Spectroscopy of Cluster Evolution with Subaru;][]{Kodama05}. 
Detection limits are listed in Table \ref{tab:allobs}. In addition
{\it K}-band imaging to a 3-$\sigma$ depth of 20.1 mag over 0.12-deg$^2$ centred on 
\ms\ was obtained from the Wide-Field Infrared Camera 
\citep[WIRC;][]{Wilson03} on the 
Palomar Hale telescope. Standard reduction methods were employed and 
are detailed in G. Smith et al. (in prep.). 
These data were used in the previous analyses of
\citet{Geach06} and \citet{Moran07a, Moran07b}.

\subsubsection{\Spitzer\ IRAC imaging}
The \Spitzer\ InfraRed Array Camera (IRAC; Fazio et al. 2004) observations of 
the $20'\times 20'$ field centered on the MS0451 field were obtained as a part 
of the Cycle 5 General Observer (program 50610, PI: M. Yun) on 2009 March 18.  
Each tile was observed for a total of 1500 seconds in each of the four 
IRAC bands (3.6, 4.5, 5.8, and 8.0 $\mu$m) in full array mode with a 15 position 
dither pattern with 100 second exposures at each position.  The new data are 
combined with the archival IRAC data (program 83) to produce 
the final mosaic images using {\sc Cluster Grinder}
\citep{Guthermuth09}, 
which is an {\sc idl} software package that utilizes standard Basic Calibrated Data (BCD) 
products from the Spitzer Science Centre's standard data pipeline.  The angular 
resolution of the final mosaic ranges between 2.0 to 2.5 arcsec depending 
on the observing band; the limiting depths are given in Table \ref{tab:allobs}.

\subsubsection{Cataloguing}
To estimate redshifts photometrically we require that the
photometry in all filters samples the same emission from the source so an accurate 
spectral energy distribution (SED) can be built. 
To meet this requirement we degrade all the optical images to match the worst 
seeing -- the ({\it V}-band) in which the FWHM is 1.54 arcsec. 
Each seeing convolved image is astrometrically matched to the USNO catalogue with 
a set of unsaturated and unblended stars spread across the frames. 
We use {\sc SExtractor} \citep{Bertin96} on the {\it R}-band image to detect 
objects with a minimum of ten adjacent 0.2-arcsec pixels at least 
1.5 $\sigma$ above the 
background to provide a source list and then use the {\sc apphot} routine in 
{\sc iraf} to extract 3-arcsec diameter aperture photometry at these positions in 
each convolved image. These measurements are then aperture corrected 
assuming a point source, to yield total magnitudes.
We report 3-$\sigma$ detection limits in Table \ref{tab:allobs}.

The resolution of the IRAC images is significantly lower than the optical 
so for source extraction we consider these data separately, although an equivalent
procedure is followed. Sources are 
detected on the 8\,$\umu$m image with {\sc SExtractor} \citep{Bertin96} and
are required to have a minimum of 4 adjacent 0.9-arcsec pixels at 
least 2 $\sigma$ above the background. The 8\,$\umu$m band is chosen so that
we can use the IRAC colours to constrain counterparts for otherwise unidentified
SMGs \citep[\S\ref{sec:iracids} and][]{Yun08}.
This source list provides 
positions for the {\sc apphot} routine in IRAF to extract 3.8-arcsec diameter fluxes. 
We also ensure that all SMGs which are identified in Submillimeter Array (SMA), 
radio or 24 $\umu$m data and have IRAC counterparts in {\it any} bands are extracted, 
removing the requirement for an 8\,$\umu$m detection.
The extracted fluxes are corrected for the aperture losses, employing the
factors calculated by the SWIRE team \citep{Surace05}, 
resulting in total magnitudes, which can be directly compared to our optical
catalogue.

\section{Analysis and Results}
\label{sec:analysis}

\begin{figure}
\includegraphics[width=8.5cm]{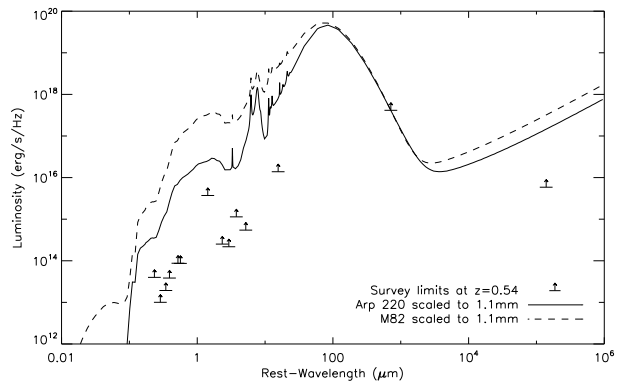}
\caption{
Catalogue detection limits (see Table \ref{tab:allobs}) in each filter for a $z=0.54$ galaxy overlayed with Arp 220 and M82 SEDs 
\citep{Silva98} scaled to our $3.5\sigma$ 1.1 mm catalogue limit. 
We expect to easily detect counterparts at this redshift suggesting that 
unidentified SMGs are most likely to be high redshift background galaxies and not cluster members.}
\label{fig:catlims}
\end{figure}

This study aims to identify 1.1-mm detected ULIRGs in the $z=0.54$ galaxy cluster \ms.
A priori, based on the space density of ULIRGs at $z\sim0.5$ \citep{LeFloch05}, and the 
overdensity of LIRGs in clusters at this redshift \citep{Geach06}, we expect to find,
at most, only a couple of cluster members in our survey.
The large background population coupled with this small number of expected members 
makes this study challenging. Therefore, we focus on techniques
to identify sources and confirm cluster membership. Similarly to other SMG studies, 
a fraction of our sources are unidentified in the radio, mid-IR or optically. However, 
based on an Arp 220 SED, cluster members which are detectable above our 1.1-mm flux limit 
are also expected to be brighter than the 
catalogue limits in all other bands (Fig. \ref{fig:catlims}).
Therefore, any unidentified sources are likely to be part of the background population.

\subsection{AzTEC catalogue}
\label{sec:azsource}

\begin{table}
\begin{minipage}{85mm} 
\caption{AzTEC galaxies in the \ms\ field in order of decreasing S/N. Source names correspond to AzTEC positions in the J2000 epoch. 
SMGs with robust multiwavelength counterparts (\S\ref{sec:allids} and Table \ref{tab:counterparts}) are shown in bold.} 
\setlength{\tabcolsep}{0.8mm}
\begin{tabular}{llcccc}
\hline
& Source & SNR & Flux & Noise & Deboosted flux\\
& & & (mJy) & (mJy) & (mJy) \\
\hline
      {\bf 1} & {\bf MMJ\,045438.96-030739.8}$^a$ & {\bf 6.5} & {\bf 8.3} & {\bf 1.3} & ${\bf 6.4^{+1.3}_{-1.4}}$\\
      {\bf 2} & {\bf MMJ\,045447.55-030018.8}$^a$ & {\bf 5.8} & {\bf 6.2} & {\bf 1.1} & ${\bf 4.7^{+1.1}_{-1.1}}$\\
      {\bf 3} & {\bf MMJ\,045433.57-025204.0}$^a$ & {\bf 5.7} & {\bf 7.7} & {\bf 1.4} & ${\bf 5.4^{+1.4}_{-1.5}}$\\
      {\bf 4} & {\bf MMJ\,045431.56-025957.8}$^b$ & {\bf 5.1} & {\bf 5.2} & {\bf 1.0} & ${\bf 3.7^{+1.0}_{-1.1}}$\\
      {\bf 5} & {\bf MMJ\,045421.55-030109.9} & {\bf 5.1} & {\bf 5.1} & {\bf 1.0} & ${\bf 3.6^{+1.0}_{-1.2}}$\\
      {\bf 6} & {\bf MMJ\,045417.49-030306.6} & {\bf 4.9} & {\bf 5.0} & {\bf 1.0} & ${\bf 3.5^{+1.1}_{-1.2}}$\\
      {\bf 7} & {\bf MMJ\,045413.35-031204.2}$^b$ & {\bf 4.6} & {\bf 6.3} & {\bf 1.4} & ${\bf 3.6^{+1.4}_{-2.0}}$\\
      {\bf 8} & {\bf MMJ\,045412.72-030043.7} & {\bf 4.5} & {\bf 4.5} & {\bf 1.0} & ${\bf 2.9^{+1.0}_{-1.3}}$\\
      {\bf 9} & {\bf MMJ\,045345.31-030552.2} & {\bf 4.4} & {\bf 4.6} & {\bf 1.0} & ${\bf 2.8^{+1.1}_{-1.4}}$\\
     {\bf 10} & {\bf MMJ\,045407.14-030033.9} & {\bf 4.3} & {\bf 4.3} & {\bf 1.0} & ${\bf 2.6^{+1.0}_{-1.4}}$\\
      11 & MMJ\,045358.12-025233.3 & 4.3 & 4.3 & 1.0 & $2.6^{+1.0}_{-1.5}$\\
      12 & MMJ\,045426.76-025806.5 & 4.2 & 4.2 & 1.0 & $2.6^{+1.0}_{-1.4}$\\
      13 & MMJ\,045356.09-025031.4 & 4.2 & 5.4 & 1.3 & $2.6^{+1.3}_{-2.2}$\\
      14 & MMJ\,045424.53-030331.7 & 4.1 & 4.2 & 1.0 & $2.4^{+1.0}_{-1.5}$\\
     {\bf 15} & {\bf MMJ\,045328.86-030243.3} & {\bf 4.1} & {\bf 5.0} & {\bf 1.2} & ${\bf 2.5^{+1.2}_{-2.0}}$\\
      16 & MMJ\,045354.64-030004.0 & 4.0 & 4.0 & 1.0 & $2.1^{+1.0}_{-1.6}$\\
     {\bf 17} & {\bf MMJ\,045431.35-025645.8} & {\bf 4.0} & {\bf 4.0} & {\bf 1.0} & ${\bf 2.3^{+1.0}_{-1.4}}$\\
     {\bf 18} & {\bf MMJ\,045411.57-030307.5} & {\bf 4.0} & {\bf 4.1} & {\bf 1.0} & ${\bf 2.3^{+1.0}_{-1.6}}$\\
      19 & MMJ\,045415.53-025125.0 & 3.9 & 4.1 & 1.0 & $2.3^{+1.1}_{-1.6}$\\
      20 & MMJ\,045345.88-030440.0 & 3.9 & 4.0 & 1.0 & $2.2^{+1.0}_{-1.6}$\\
      21 & MMJ\,045358.49-025601.4 & 3.9 & 3.8 & 1.0 & $2.2^{+1.0}_{-1.5}$\\
      22 & MMJ\,045403.10-025006.8 & 3.9 & 4.7 & 1.2 & $2.2^{+1.3}_{-2.0}$\\
      23 & MMJ\,045357.46-030237.1 & 3.9 & 4.0 & 1.0 & $2.1^{+1.0}_{-1.6}$\\
      24 & MMJ\,045420.10-030658.2 & 3.8 & 4.0 & 1.0 & $2.2^{+1.1}_{-1.6}$\\
      25 & MMJ\,045422.17-030445.8 & 3.8 & 3.9 & 1.0 & $2.0^{+1.0}_{-1.7}$\\
     {\bf 26} & {\bf MMJ\,045349.69-025807.1} & {\bf 3.8} & {\bf 3.8} & {\bf 1.0} & ${\bf 2.0^{+1.0}_{-1.6}}$\\
     {\bf 27} & {\bf MMJ\,045421.17-030740.2} & {\bf 3.8} & {\bf 3.9} & {\bf 1.0} & ${\bf 1.9^{+1.1}_{-1.7}}$\\
      28 & MMJ\,045345.06-025722.6 & 3.7 & 3.7 & 1.0 & $1.9^{+1.0}_{-1.6}$\\
      29 & MMJ\,045442.54-025455.0 & 3.7 & 4.5 & 1.2 & $1.6^{+1.4}_{-1.6}$\\
      30 & MMJ\,045444.78-030030.9 & 3.7 & 3.8 & 1.0 & $1.9^{+1.1}_{-1.6}$\\
      31 & MMJ\,045411.73-025712.7 & 3.7 & 3.6 & 1.0 & $1.8^{+1.0}_{-1.6}$\\
      32 & MMJ\,045411.17-031019.2 & 3.6 & 4.1 & 1.1 & $1.8^{+1.2}_{-1.7}$\\
      33 & MMJ\,045352.72-030130.9 & 3.6 & 3.7 & 1.0 & $2.0^{+1.0}_{-1.6}$\\
      34 & MMJ\,045454.20-025919.3 & 3.6 & 4.6 & 1.3 & $1.5^{+1.5}_{-1.5}$\\
      35 & MMJ\,045359.27-030327.4 & 3.5 & 3.6 & 1.0 & $1.8^{+1.0}_{-1.6}$\\
      36 & MMJ\,045340.09-030334.2 & 3.5 & 3.6 & 1.0 & $1.7^{+1.0}_{-1.6}$\\
\hline
\end{tabular}
\label{tab:smgs}
\\ $^a$ These SMGs were observed with the SMA and detected (\S\ref{sec:sma}).
\\ $^b$ These SMGs were observed with the SMA but not formally detected (\S\ref{sec:sma}).
\end{minipage} 
\end{table}

\begin{figure}
\includegraphics[width=80mm]{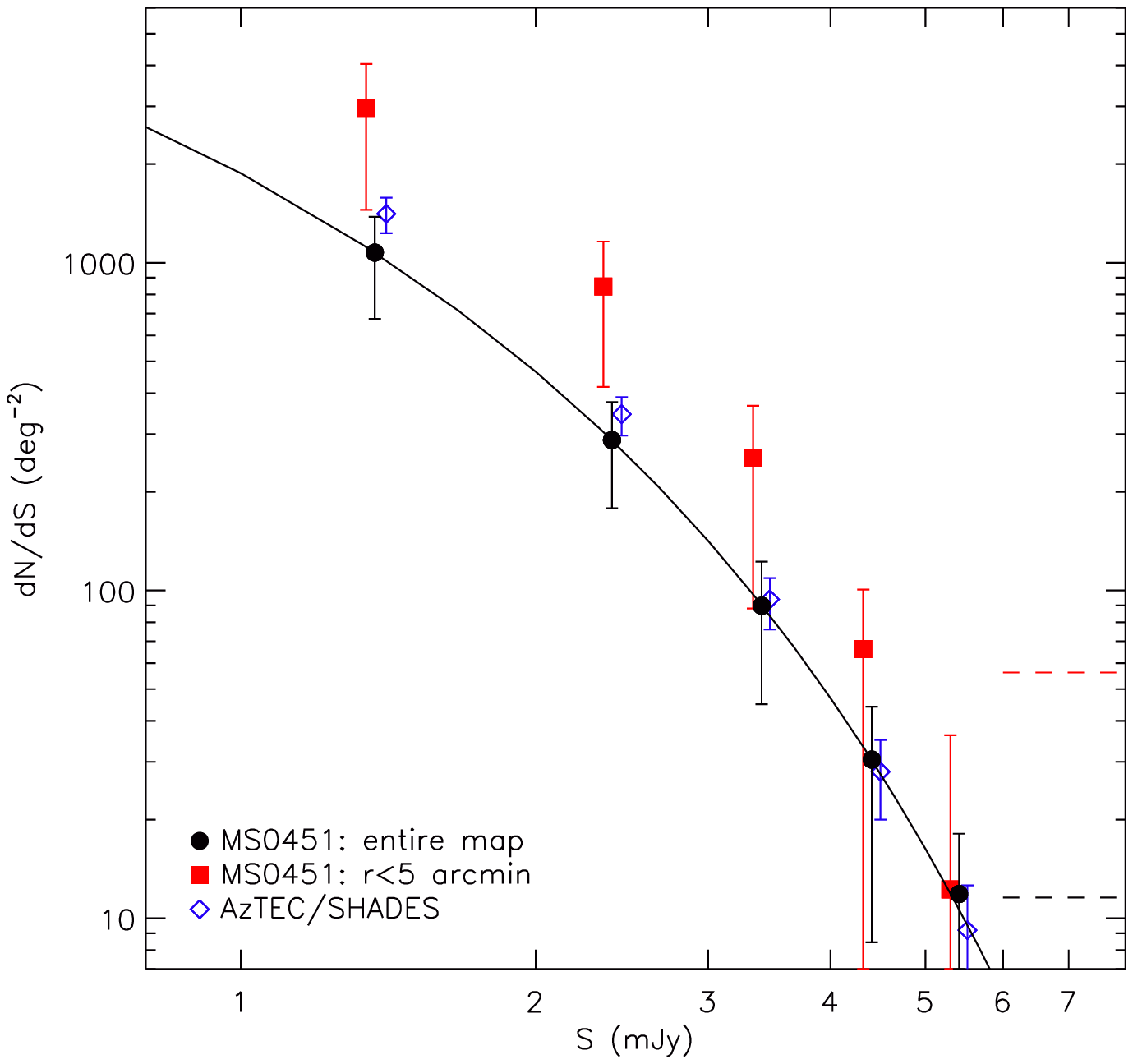}
\caption{Differential number counts using deboosted fluxes from the  AzTEC survey of \ms\ compared to 
those from the AzTEC/SHADES survey \citep{Austermann09b}. 
We see that across the whole area of our survey
there is no strong overdensity of 1.1 mm sources in \ms\ compared to the AzTEC/SHADES 
blank-field survey; hence it is apparent that the majority, but not necessarily the entirety, of the 
millimetre sources in the cluster field are background galaxies.
However, in the central 5 arcmin radius of the map ($\sim1.9$ Mpc at $z=0.54$) there appears to be a slight overdensity 
of sources. This could be due to a combination of the Sunyaev-Zel'dovich
effect, gravitational lensing of background sources by \ms, and 
potentially a contribution from 1.1 mm cluster members.
The solid line is the Schechter function fit to the number counts in the entire
AzTEC map, and the dotted lines represent the survey limits - the lower and upper lines represent the whole
AzTEC \ms\ field and the central 5 arcmin radius, respectively.
For clarity number counts in the central 5 arcmin radius of \ms\ and the AzTEC/SHADES survey are offset slightly in flux.
}
\label{fig:numbercounts}
\end{figure}

Millimetre sources are identified as local maxima with S/N$\ge3.5$ and are selected in the roughly uniform 
noise ($\sigma\sim1.1$ mJy, $\sim$360-arcmin$^2$) region of the AzTEC signal-to-noise map 
(with weights of at least half of the maximum). 
In Table \ref{tab:smgs} we present these 36 detections in order of decreasing S/N. 
Measured source fluxes and their errors are given by the peak pixel value and corresponding 
noise value respectively; deboosted fluxes are calculated following the method of \citet{Perera08}.
We determine the positions of sources on the sub-pixel level by calculating the centroid after
weighting pixels close to the brightest pixel according to the square of their fluxes.
Fig. \ref{fig:smgmap} shows the Subaru {\it R}-band image with AzTEC contours of 
constant S/N and the detected 1.1-mm sources marked. 
Based on the analyses of AzTEC data in \citet{Scott08} and \citet{Perera08} we expect that 3--4 (8--10\%) 
of the sources in our $3.5\sigma$ catalogue are false detections with 
only $\sim0.5$ and 0 amongst those with S/N$\geq 4.5$ and $5$ respectively.

To examine whether there is an overdensity of AzTEC galaxies in \ms\ 
compared to similar field surveys -- potentially indicating a significant number of obscured 
ULIRG cluster members -- we calculate the number counts in \ms\ compared to the AzTEC 
blank-field survey of the SHADES fields \citep{Austermann09b}.
The source number counts and corresponding errors were calculated using the same bootstrap 
sampling methods as described in \citet{Austermann09b}, employing the 
AzTEC/SHADES best fit Schechter parameters as a prior. 
Due to the effects of flux boosting, our catalogue, which is limited at an apparent S/N ratio of 3.5,
corresponding to a typical apparent flux of 3.6 mJy, actually contains sources fainter than this limit, 
allowing us to statistically constrain counts fainter than 3.6 mJy.
We also calculate number counts in the central 5 arcmin radius of the AzTEC \ms\ map, by
trimming the map to this area and repeating the calculations with the sky model, deboosting, and
completeness estimates based on the full maps, since these are not expected to change across
the field. The number counts for the whole \ms\ survey area, and the central 5 arcmin  
(which represents a physical scale of $\sim1.9$ Mpc at $z=0.54$),
compared to the AzTEC/SHADES survey are presented in Fig. \ref{fig:numbercounts}.

Across the whole field \ms\ does not exhibit a source excess at 1.1 mm
compared to a blank field. This suggests that, as expected, there is not
a dominant population of luminous obscured star-forming galaxies in \ms.
Nevertheless, it is possible that a small number of the sources are still cluster members. 
Indeed, our number counts analysis (Fig. \ref{fig:numbercounts}) suggests there may be a small 
overdensity of sources in the central 5-arcmin radius region of the AzTEC map.
Based on integral number counts and their errors at 1.1 mm down to 1 mJy, derived from bootstrap sampling the 
Posterior Flux Densities of sources (as described in detail in \citet{Austermann09b}) 
a tentative overdensity is apparent at the $\sim1.6\sigma$ level.
If real this overdensity could arise from a combination of the Sunyaev-Zel'dovich
effect, the gravitational lensing of background sources by the cluster, and potentially 
1.1 mm cluster members. To discover whether any of this excess is caused by 1.1-mm bright 
cluster galaxies we must first accurately locate the counterparts to the 1.1-mm emission.

\subsection{Identifying counterparts to AzTEC sources}
\label{sec:allids}
The large AzTEC beam (18-arcsec FWHM on the JCMT), coupled with the high spatial density 
of optical sources makes identification difficult unless precise positions are determined for the SMGs. 
The observed mm/sub-mm emission from SMGs represents rest-frame far-IR 
emission from dust-reprocessed starlight or AGN activity. Therefore, by exploiting 
the far-IR--radio correlation \citep[e.g.][]{Condon92,Garrett02} and the low spatial density of radio sources,
it is possible to employ deep ($\sigma\sim10\, \rm{\umu Jy}$), 
high-resolution, interferometric radio observations to determine accurate source positions 
for $\sim$60--80 per cent of SMGs \citep[e.g.][]{Ivison02,Ivison05,Ivison07,Chapin09}.
Similarly, 24-$\umu$m observations can be used for identification, a method which has proved useful in confirming tentative radio 
counterparts, or providing positions for a small number of the radio undetected SMGs \citep[e.g.][]{Ivison04,Pope06,Ivison07,Chapin09}.
In addition, mid-IR colours have been used to isolate potential cluster SMGs -- 
although this emission does not track far-IR output and so it is an indirect indicator of the
luminous far-IR source \citep{Ashby06, Pope06, Yun08}.

An alternative approach is submillimetre interferometry \citep[e.g.][]{Iono06,Younger07,Cowie09},
which is preferable to these traditional radio and mid-IR identification methods because it samples 
the same part of the Spectral Energy Distribution (SED) as our AzTEC imaging.
However, sources generally require individual observations due to the small field-of-views 
of such instruments, commanding prohibitive exposure times to accurately locate all 
sources in a catalogue. \\

We employ all three methods in this work, using millimetre 
interferometry where available, and radio and mid-IR imaging as an alternative.
The details of each of these identification methods is discussed in the following sections.
In total we identify 18 AzTEC counterparts (50\% of the sample; Table \ref{tab:counterparts}), 
of which 14 (39\% of the total) also 
have optical counterparts (Table \ref{tab:optprops}). In Fig. \ref{fig:catlims} we 
demonstrate that at $z\sim 0.5$ our multiwavelength observations are deep enough to detect 
galaxies above our AzTEC flux limit and hence provide counterparts, assuming SEDs typical of Arp 220 or M82. 
Therefore, our unidentified sample is not expected to contain cluster members.
Of the optically-unidentified SMGs, one lies out of the field of the Subaru, CFHT and IRAC observations, 
two are contaminated by nearby saturated stars and one is fainter than 
our detection limits at all wavelengths. 
We discuss the AzTEC galaxies and corresponding counterparts on a source by 
source basis in Appendix \ref{sourcenotes}.

\subsubsection{SMA detections}
\label{sec:sma}
The five brightest SMGs presented here were observed at $890\ \umu$m with the 
SMA in October and November 2007. 
On-source integration times of 5--6 hours were employed, yielding maps with 
$\sigma_{\rm 890\mu m}\sim1.6$ mJy. 
The SMA configuration resulted in a beam of $\sim 3$ arcsec. 
These data are discussed in detail in Wardlow et al. (in prep.).

Of the five sources observed with the SMA MMJ\,045438.96, MMJ\,045447.55 and 
MMJ\,045433.57 were detected with fluxes of 8.4, 5.5 and 8.3 mJy respectively. The
remaining two targets, MMJ\,045431.56 and MMJ\,045413.35, were undetected, but 
have AzTEC 1.1 mm detections with ${\rm SNR}=5.1$ and $4.6$ respectively.
Unlike this \ms\ study, previous SMA observations of AzTEC galaxies in the COSMOS 
field detected all seven targets \citep{Younger07}. However, 
the faintest of the COSMOS SMGs has deboosted 1.1 mm flux of 5.2 mJy, compared to 
3.7 and 3.6 mJy for MMJ\,045431.56 and MMJ\,045413.35 respectively.
Therefore, based on the deboosted fluxes it is possible that MMJ\,045431.56 and 
MMJ\,045413.35 are too faint to be detected in our observations, although
there are examples of bright SMGs that are undetected with the SMA \citep[e.g.][]{Matsuda07}.
Alternatively, the lack of SMA counterparts could suggest that either
the $890 \umu$m to 1.1 mm flux ratios of these sources are lower than expected 
(i.e. the dust is colder, or they are higher redshift than spectroscopically identified SMGs), 
that these galaxies exhibit extended far-IR emission on scales
$\gg2$ arcsec ($\gg16$ kpc at $z=2$; for example, from a merger-induced starburst), or that 
the AzTEC beam contains a blend of multiple SMGs. 
Our radio and SMA observations have similar resolutions, therefore, extended or multiple radio 
counterparts can indicate the extended or multiple nature of (sub)millimetre emission. 
Based on the radio emission it is likely that MMJ\,045431.56 is composed of multiple 
(sub)millimetre sources, as discussed in Appendix \ref{sourcenotes}. However, MMJ\,045413.35 
is more likely to be a single resolved (sub)millimetre source, with a lower 890-to-1100 $\umu$m ratio
than expected.
These issues are discussed further in Wardlow et al. (in prep.). 
We use the positions of the three SMA detected SMGs in the following analysis.

We note that the proposed radio identifications of MMJ\,045438.96 and MMJ\,045433.57 
are coincident with their SMA positions. 
The other SMA detected galaxy, MMJ\,045447.55, lies $\sim 9$ arcsec from an elliptical galaxy with
bright radio and mid-IR emission, which, due to the high fluxes is formally a
`robust' identification, but is unlikely to be the true source of the millimetre emission.
MMJ\,045431.56 and MMJ\,045413.35 which were targeted but not detected with the 
SMA are identified through radio and mid-IR counterparts respectively.

\begin{landscape}
\begin{table}
\begin{minipage}{240mm} 
\caption{Radio and 24\,$\umu$m counterparts of AzTEC galaxies in \ms. All matches within 10-arcsec are 
listed, those secure identifications at 1.4 GHz or 24 $\umu$m with $P\leq0.05$ are shown in bold and tentative associations ($0.05<P\le 0.10$) are also presented.
Images and a discussion of each source are presented in Appendix \ref{sourcenotes}.}
\setlength{\tabcolsep}{1.5 mm}
\begin{tabular}{llccccccccccccc}
\hline
  \multicolumn{1}{c}{} &
  \multicolumn{1}{c}{Source} &
  \multicolumn{2}{c}{1.4GHz position\footnote{1.4 GHz positions have typical uncertainties of $\sim0.6$-arcsec.}} &
  \multicolumn{1}{c}{$S_{{\rm 1.4GHz}}$\footnote{1.4 GHz fluxes have typical errors of 17 $\umu$Jy.}} &
  \multicolumn{1}{c}{1.1mm--1.4GHz} &
  \multicolumn{1}{c}{$ P_{\rm 1.4GHz}$} &
  \multicolumn{2}{c}{24$\umu$m position\footnote{24 $\umu$m positions have typical uncertainties of $\sim1.8$-arcsec.}} &
  \multicolumn{1}{c}{$S_{{\rm 24\umu m}}$\footnote{24 $\umu$m fluxes have typical errors of 40 $\umu$Jy.} }&
  \multicolumn{1}{c}{1.1mm--24$\umu$m} &
  \multicolumn{1}{c}{$P_{\rm 24\umu m}$} &
  \multicolumn{1}{c}{24$\umu$m--1.4GHz\footnote{24$\umu$m--1.4GHz separations consistent with the same counterpart are italicised.}} \\
&  &            RA & Dec        &            & Separation & &  RA & Dec                   &            & Separation   & & Separation \\
&  &\multicolumn{2}{c}{(J2000)} & ($\umu$Jy) & (arcsec)   & & \multicolumn{2}{c}{(J2000)} & ($\umu$Jy) & (arcsec)     & & (arcsec) \\
\hline
\bf 1  &\bf MMJ\,045438.96\footnote{The radio and mid-IR positions agree with the SMA position (\S\ref{sec:sma})}  &\bf 04$^{\rm\bf h}$54$^{\rm\bf m}$39$\fs$01 &\bf $\bf-03\degr$07$\arcmin$38$\farcs$5 &\bf 98.0 &\bf 1.51 & \bf 0.003 &\bf 04$^{\rm\bf h}$54$^{\rm\bf m}$38$\fs$97 &\bf $\bf-03\degr$07$\arcmin$38$\farcs$0 &\bf 157 &     \bf   1.3 &\bf 0.023 & \it 0.79  \\
\bf 2 &\bf MMJ\,045447.55\footnote{MMJ\,045447.55 is identified from SMA observations (Wardlow et al., 2009b), but a bright elliptical galaxy $\sim9.6$ arcsec from the AzTEC position is also a formal radio and mid-IR identification. We consider the radio and mid-IR galaxy to be a chance association -- at $P=0.05$ we expect 1--2 chance associations in our catalogue. }&\bf 04$^{\rm\bf h}$54$^{\rm\bf m}$47$\fs$46 &\bf $\bf-03\degr$00$\arcmin$19$\farcs$2 &  $<51$ &- & -  & -   &  -                               &          $<120$                       &     -        & -    & -           &     \\
  &                & 04$^{\rm h}$54$^{\rm m}$48$\fs$14 & $-03\degr$00$\arcmin$14$\farcs$6 &  $101$ & 9.78 & 0.032 & 04$^{\rm h}$54$^{\rm m}$48$\fs$12  & $-03\degr$00$\arcmin$14$\farcs$3 & 3396   & 9.56 & 0.009  & \it 0.46  &     \\
\bf 3  &\bf MMJ\,045433.57$^f$  &\bf 04$^{\rm\bf h}$54$^{\rm\bf m}$33$\fs$55 &\bf $\bf-02\degr$52$\arcmin$04$\farcs$6 &\bf 93.5 &\bf 0.71 & \bf 0.0008  &\bf 04$^{\rm\bf h}$54$^{\rm\bf m}$33$\fs$51 &\bf $\bf-02\degr$52$\arcmin$06$\farcs$8 &\bf 392 &\bf        3.00 &\bf 0.020    & \it 2.28  \\
\bf 4  &\bf MMJ\,045431.56\footnote{There is an additonal 176-$\umu$Jy 24-$\umu$m source, with $P=0.17$, coincident with the radio counterpart of MMJ\,045431.56. It is considered related to the AzTEC detection due to the radio identification.}   &\bf 04$^{\rm\bf h}$54$^{\rm\bf m}$31$\fs$69 &\bf $\bf-03\degr$00$\arcmin$07$\farcs$1 &\bf 77.8 &\bf 9.50 & \bf 0.035   & 04$^{\rm h}$54$^{\rm m}$31$\fs$13          & $-02\degr$59$\arcmin$51$\farcs$6    & 485    &       8.89     & 0.076       & 17.62    \\
\bf 5  &\bf MMJ\,045421.55      &\bf 04$^{\rm\bf h}$54$^{\rm\bf m}$21$\fs$66 &\bf $\bf-03\degr$01$\arcmin$08$\farcs$5 &\bf 84.3  &\bf 2.21 & \bf 0.006   &\bf 04$^{\rm\bf h}$54$^{\rm\bf m}$21$\fs$40 &\bf $\bf-03\degr$01$\arcmin$07$\farcs$5 &\bf 209 & \bf       3.29 &\bf 0.046    & 4.09  \\
\bf 6  &\bf MMJ\,045417.49      &\bf 04$^{\rm\bf h}$54$^{\rm\bf m}$17$\fs$27 &\bf $\bf-03\degr$03$\arcmin$03$\farcs$2 &\bf 189   &\bf 4.74 & \bf 0.010   &     04$^{\rm h}$54$^{\rm m}$17$\fs$24 &    $-03\degr$03$\arcmin$03$\farcs$7 &    217 &           4.76 &    0.074    & \it 0.58  \\
\bf 7  &\bf  MMJ\,045413.35      &\bf 04$^{\rm\bf h}$54$^{\rm\bf m}$13$\fs$36 &\bf $\bf-03\degr$11$\arcmin$58$\farcs$8 &\bf 78.4 &\bf 5.44 & \bf 0.021   &     04$^{\rm  f h}$54$^{\rm    m}$13$\fs$38 &    $-03\degr$11$\arcmin$58$\farcs$9 &    333 &           5.31 &    0.056    &  \it 0.29   \\
\bf 8  &\bf MMJ\,045412.72      &\bf 04$^{\rm\bf h}$54$^{\rm\bf m}$12$\fs$79 &\bf $\bf-03\degr$00$\arcmin$43$\farcs$9 & \bf 58.3 &\bf 1.03 & \bf 0.002   &  -  &           -                      &                   $<120$              &      -       &  -   &     -       &     \\
\bf 9  &\bf MMJ\,045345.31      & \bf04$^{\rm\bf h}$53$^{\rm\bf m}$45$\fs$29 &\bf $\bf-03\degr$05$\arcmin$53$\farcs$5 &\bf 92.3  &\bf 1.30 & \bf 0.002   &\bf 04$^{\rm\bf h}$53$^{\rm\bf m}$45$\fs$36 &\bf $\bf-03\degr$05$\arcmin$52$\farcs$8 &\bf 1393& \bf       0.86 &\bf 0.0004   &\it 1.24  \\
   &                      &\bf 04$^{\rm\bf h}$53$^{\rm\bf m}$45$\fs$07 &\bf $\bf-03\degr$05$\arcmin$56$\farcs$6 &\bf 76.9  &\bf 5.62 &\bf 0.022  &                                  &                                              &           &            &            & 5.71  \\
\bf 10 &\bf MMJ\,045407.14 &\bf 04$^{\rm\bf h}$54$^{\rm\bf m}$07$\fs$08 &\bf $\bf-03\degr$00$\arcmin$37$\farcs$2 &\bf 60.9  &\bf 3.37 &\bf 0.013   &\bf 04$^{\rm\bf h}$54$^{\rm\bf m}$07$\fs$07 &\bf $\bf-03\degr$00$\arcmin$37$\farcs$5 &\bf 1120&\bf        3.76  &\bf 0.007  &\it 0.39  \\
12 &    MMJ\,045426.76\footnote{MMJ\,045426.76 and MMJ\,045442.54 have no radio or mid-IR counterparts but are identified on the basis of their IRAC colours \citep{Yun08}.}& 04$^{\rm h}$54$^{\rm m}$26$\fs$86 & $-02\degr$58$\arcmin$08$\farcs$0 &  $<51$ &- & -  & -   &  -                               &          $<120$                       &     -        & -    & -           &     \\
\bf 15 &\bf MMJ\,045328.86\footnote{There is a $\sim3\sigma$ radio peak coincident with the 24-$\umu$m counterpart.} &        -                             &                             -       &   $\sim 51$   &  -   &    -              &\bf 04$^{\rm\bf h}$53$^{\rm\bf m}$28$\fs$83 &\bf $\bf-03\degr$02$\arcmin$45$\farcs$9 &\bf 150 &\bf        2.54  &\bf 0.040  &-      \\
\bf 17 &\bf MMJ\,045431.35    &\bf 04$^{\rm\bf h}$54$^{\rm\bf m}$31$\fs$71 &\bf $\bf-02\degr$56$\arcmin$45$\farcs$0 &\bf 89.3 &\bf 5.41 &\bf 0.019 &                         -         &      -                                       &   $<120$    &          -      &    -   &      -\\
\bf 18 &\bf MMJ\,045411.57$^j$ &        -                             &                             -       &   $\sim 51$   &  -   &    -              &\bf 04$^{\rm\bf h}$54$^{\rm\bf m}$11$\fs$66 &\bf $\bf-03\degr$03$\arcmin$08$\farcs$0 &\bf 185 & \bf       1.55  &\bf 0.016  & -     \\
\bf 26 &\bf MMJ\,045349.69     &        -                             &                             -       &   $   <51$   &   -  &    -              &\bf 04$^{\rm\bf h}$53$^{\rm\bf m}$49$\fs$80 &\bf $\bf-02\degr$58$\arcmin$13$\farcs$9 &\bf 530 & \bf      7.03   &\bf 0.049  &  -    \\
\bf 27 &\bf MMJ\,045421.17$^j$ &        -                             &                             -       &   $\sim 51$   &  -   &   -               &\bf 04$^{\rm\bf h}$54$^{\rm\bf m}$21$\fs$14 &\bf $\bf-03\degr$07$\arcmin$46$\farcs$1 &\bf 464 &\bf        5.92  &\bf 0.045  &-      \\
28 &    MMJ\,045345.06$^j$ &        -                             &                             -       &   $\sim 51$   &  -   &   -               &    04$^{\rm    h}$53$^{\rm    m}$45$\fs$33 &    $-02\degr$57$\arcmin$18$\farcs$1 &    226 &           6.00  &    0.095  &-      \\
29 &    MMJ\,045442.54$^i$ & 04$^{\rm h}$54$^{\rm m}$42$\fs$62 & $-02\degr$54$\arcmin$48$\farcs$2 &  $<51$ &- & -  & -   &  -                               &         $<120$                       &     -        & -    & -           &     \\
\hline
\end{tabular}

\label{tab:counterparts}
\end{minipage} 
\end{table}
\end{landscape}

\subsubsection{Radio and mid-IR identifications}
\label{sec:ids}

In this work we use both 1.4-GHz VLA and 24-$\umu$m \textit{Spitzer} MIPS observations 
to locate the AzTEC galaxies in our sample. 
We identify SMG counterparts independently at radio and mid-IR wavelengths 
before comparing these for each AzTEC source.
In our maps the surface density of radio galaxies is lower than the mid-IR, 
the positional accuracy is greater ($\sim 0.6$ arcsec compared to $\sim 1.8$ arcsec), 
and the link from the radio emission to the far-IR is tighter.
Therefore, if the 24\,$\umu$m and 1.4 GHz identifications disagree we consider 
the radio position as the more reliable counterpart.

To ensure that no genuine associations are missed we search up to 10 arcsec from each AzTEC position. 
This corresponds to a $\sim3\sigma$ search radius for the lowest signal-to-noise 
ratio (SNR) sources, where $\sigma$ is the error on AzTEC position, 
given by $0.6 \rm{FWHM} \rm{(SNR)}^{-1}$, and the FWHM is of the instrument beam \citep{Ivison07}.
We reject counterparts with more than 10 per cent probability of being chance associations using the 
P-statistic of \citet{Downes86} \citep[see also][]{Pope06, Ivison07}. Counterparts found with $P<0.05$ are considered robust and
those with $0.05<P<0.1$ tentative. We present all identifications in Table \ref{tab:counterparts}.

\subsubsection{IRAC identifications}
\label{sec:iracids}

\citet{Yun08} studied SMGs securely identified with SMA, radio, or 24\,$\umu$m data
and found that they typically have redder IRAC colours than the submillimetre-faint foreground
galaxy population. They proposed a selection criteria for SMG counterparts, based on
IRAC colours and found that the SMA detected galaxies
in their sample, whether radio and mid-IR identified or not, were all recovered by 
this method. 
We verify that this selection criteria for our SMA, radio and 24 $\umu$m-identified SMG counterparts
and use it to search for
counterparts to our otherwise unidentified AzTEC sources and
identify two further galaxies. 
Unlike the \citet{Yun08} SMA-detected SMGs, MMJ\,045447.55, our optically, radio and 
24$\umu$m faint, but SMA-detected SMG, is not detected in any of the IRAC wavebands
to the depth of the survey listed in Table \ref{tab:allobs}.

\subsection{Redshifts of \ms\ AzTEC galaxies}
\label{sec:photoz}

Our study seeks to identify potential millimetre sources in the cluster population of \ms.
There are spectroscopic observations of 1639 galaxies within our field, but nevertheless none of 
the AzTEC galaxies have been spectroscopically observed. 
Therefore, we use photometric methods to separate any potential AzTEC detected cluster members from `typical' $z\sim2$ SMGs. 
We first apply two simple colour tests -- the advantage of these is that it is easy 
to understand the biases in the sample -- before applying a more sophisticated photometric 
redshift analysis.
We use the $S_{24{\rm  \umu m}} / S_{1.1 {\rm mm}}$ and $S_{1.4 {\rm GHz}} / S_{1.1 {\rm mm}}$ flux ratios 
to estimate the redshifts of identified AzTEC galaxies and then also consider 
the $BzK$ selection criteria of \citet{Daddi04} to separate the higher redshift ($z>1.4$) SMGs from potential cluster members.
Finally, we test whether these conclusions are reliable by using the spectroscopic 
redshifts and multi-wavelength photometry of the SMGs in \citet{Borys05}, 
and apply these findings to our AzTEC \ms\ sources to obtain photometric redshifts for our galaxies.

\subsubsection{Simple photometric redshift analysis}
\label{sec:simplephoto}

\begin{figure}
\includegraphics[width=8.5cm]{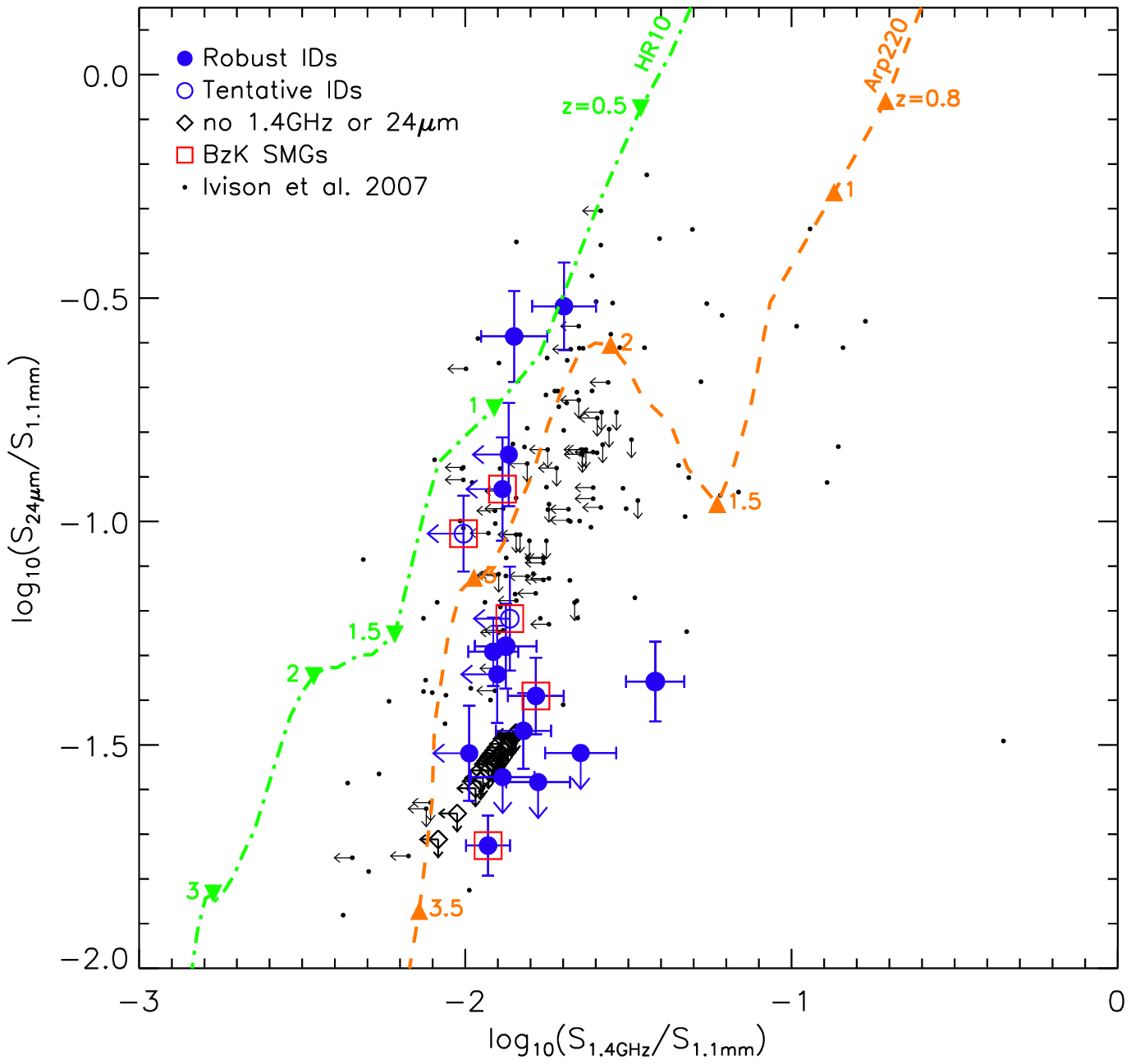}
\caption{$S_{24 {\rm \umu m}} / S_{1.1{\rm  mm}}$ versus $S_{1.4 {\rm GHz}} / S_{1.1{\rm  mm}}$ for SMGs in the \ms\ field.
We differentiate between sources with robust and tentative detections ($0.05<P\le 0.10$), 
and highlight SMGs which satisfy the $BzK$ selection criteria of \citet{Daddi04} (Fig. \ref{fig:bzk}).
Sources without both mid-IR and radio detections are plotted at the 3$\sigma$ detection limit of the respective catalogues (Table \ref{tab:allobs}).
For comparison we also plot SHADES identifications \citep{Ivison07}, converted to 
expected observed 1.1 mm fluxes as discussed in the text; error bars of this sample are omitted for clarity. 
Both sets of SMGs lie broadly within the same region of colour-colour space, providing confidence in our identifications.
Redshift tracks of the $z=1.4$ SMG HR 10 and local starburst Arp 220 (based on SEDs from \citet{Silva98}) suggest that the AzTEC sources generally have 
$1\la z\la 3.5$, in concordance with \citet{Chapman05}.
}
\label{fig:radmir}
\end{figure}

In Fig. \ref{fig:radmir} we plot $S_{24{\rm  \umu m}} / S_{1.1 {\rm mm}}$ versus 
$S_{1.4 {\rm GHz}} / S_{1.1 {\rm mm}}$ for AzTEC \ms\
SMGs. For comparison we plot 850\,$\umu$m sources from \citet{Ivison07}, extrapolated from the observed 850\,$\umu$m to the 
equivalent $1.1$mm flux assuming a power law spectrum of the form $S_{\nu} \sim \nu^{3.5}$. 
As expected both sets of SMGs lie broadly within the same region of colour-colour space, adding confidence to our detections and identifications.
The redshift tracks of the local star-forming galaxy Arp 220, and the higher redshift HR10 ($z=1.44$) (based on the SEDs of \citet{Silva98}) suggest that the 
AzTEC galaxies have redshifts of $1\la z\la 3.5$, in concordance with the SMG population \citep{Chapman05}. However, potential
degeneracy between redshift and dust temperature in the templates makes this method unreliable for identifying cluster galaxies.

\begin{figure}
\includegraphics[width=8.5cm]{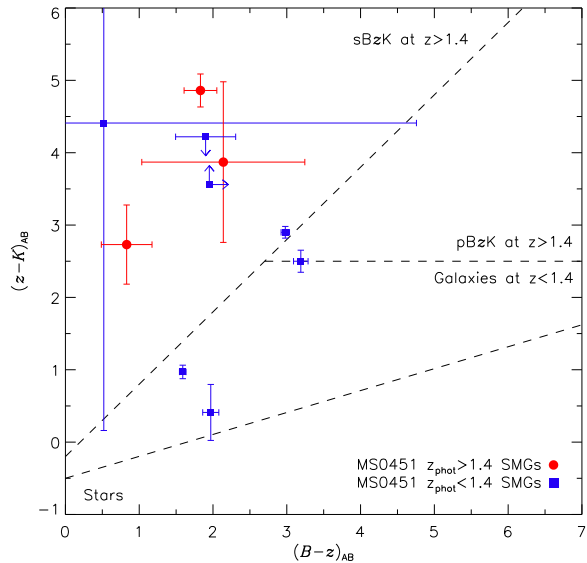}
\caption{$B-z$ vs. $z-K$ colour-colour plot of the optically identified AzTEC galaxies in \ms\ which lie within the field-of-view of our {\it B}, {\it z} and 
{\it K} observations. 
The lines of separation between passive and star-forming $z>1.4$ $BzK$ galaxies (pBzK and sBzK respectively), $z<1.4$ galaxies, and stars \citep{Daddi04} are shown.
Two SMGs occupy the $z<1.4$ region, suggesting they are potential cluster members.
We distinguish between galaxies with $z_{phot}>1.4$ and $z_{phot}<1.4$ based on photometric redshifts calculated in \S\ref{sec:aztecphotoz}. 
The $BzK$ selection and full photometric analysis broadly agree as to the
high and low redshift samples giving more confidence in our ability to isolate millimetre sources in the cluster from the dominant background population.
As expected, by this criteria most SMGs are deemed to be actively star-forming.
}
\label{fig:bzk}
\end{figure}

We next determine whether any of the SMGs are likely to lie at low redshift and hence potentially be members of \ms\ using the $BzK$ selection of 
\citet{Daddi04} to separate $z>1.4$ galaxies from those at $z<1$.
Of the $\sim70$ SMGs with spectroscopic redshifts presented in \citet{Chapman05} $\sim80\%$ 
have $z>1.4$, suggesting that the $BzK$ selection should enable us to remove the majority of 
background sources from our sample.
Fig. \ref{fig:bzk} shows the optically identified SMGs in \ms\ which are covered by our {\it B}, {\it z} and {\it K} imaging, in addition to
\citet{Daddi04} selection criteria for high redshift ($z>1.4$) galaxies; 
SMGs with $z_{phot}<1.4$ are highlighted (\S\ref{sec:aztecphotoz}).
It is clear from Fig. \ref{fig:bzk} that from a $BzK$ selection alone, at least two SMGs are low redshift and therefore potential cluster members;
several additional galaxies lie close to the border or have photometric limits which could place them in the low redshift region. 
Therefore, we next carry out a full photometric redshift analysis of the whole sample using
{\sc Hyperz} \citep{Bolzonella00} to calculate redshifts for the SMGs

\begin{landscape}
\begin{table}  
\begin{minipage}{240mm} 
\caption{Optical and near-IR photometry for the detected SMG counterparts with derived photometric redshifts.
Potential cluster members are shown in bold (\S\ref{sec:aztecphotoz}).}
\setlength{\tabcolsep}{1.2 mm}
\begin{tabular}{llcccccccccccccccccccccc}
\hline
 & Source & \it U &\it B &\it V &\it R &\it I &\it z &\it K & 3.6$\umu$m & 4.5$\umu$m & 5.8$\umu$m & 8$\umu$m& $z_{phot}$ \\
\hline
   1 & MMJ\,045438.96 &      $>25.1$   &  25.89$\pm$0.04 &  24.03$\pm$0.04 &  23.29$\pm$0.03 &  22.77$\pm$0.03 &  23.75$\pm$1.10 &  19.88$\pm$0.11 &  20.78$\pm$0.05 &  20.37$\pm$0.03 &  19.92$\pm$0.09 &  20.44$\pm$0.07  &  $3.42^{+0.07}_{-0.05}$ \\
   3 & MMJ\,045433.57 & 26.92$\pm$0.17 &  25.60$\pm$0.04 &  24.96$\pm$0.05 &  24.02$\pm$0.03 &  23.35$\pm$0.04 &        -        &  19.27$\pm$0.07 &  20.41$\pm$0.03 &  20.05$\pm$0.02 &  19.72$\pm$0.08 &  20.11$\pm$0.05  &  $2.55^{+0.06}_{-0.09}$ \\
   4 & MMJ\,045431.56 & 24.04$\pm$0.04 &  24.20$\pm$0.02 &  23.45$\pm$0.03 &  22.71$\pm$0.02 &  21.97$\pm$0.03 &  21.22$\pm$0.05 &  18.32$\pm$0.06 &  20.18$\pm$0.03 &  20.14$\pm$0.02 &  20.15$\pm$0.11 &  20.85$\pm$0.10  &  $0.86^{+0.04}_{-0.03}$ \\
   \bf 5\footnote{This is the combined photometry for C1 and C2 as discussed in \S\ref{sec:aztecphotoz}} &\bf MMJ\,045421.55 &\bf 24.92$\pm$0.08 &\bf  24.50$\pm$0.03 &\bf  23.37$\pm$0.03 &\bf  22.31$\pm$0.02 &\bf  21.44$\pm$0.03 &\bf  21.31$\pm$0.09 &\bf  18.81$\pm$0.12 &\bf  19.70$\pm$0.02 &\bf  19.96$\pm$0.02 & \bf 19.43$\pm$0.06 & \bf 20.54$\pm$0.08  &  $\bf{0.50^{+0.05}_{-0.10}}$ \\
   8 & MMJ\,045412.72 &      $>25.1$   &      $>26.6$    &     $>25.8$     &      $>25.1$    &      $>24.2$    &     $>24.2$     &      $>20.1$    &  19.48$\pm$0.01 &  19.03$\pm$0.02 &  18.61$\pm$0.03 &  19.40$\pm$0.03  &  $1.85^{+0.31}_{-0.15}$ \\
  10 & MMJ\,045407.14 & 22.83$\pm$0.03 &  22.23$\pm$0.02 &  21.69$\pm$0.02 &  21.23$\pm$0.02 &  20.80$\pm$0.02 &  20.64$\pm$0.04 &  19.67$\pm$0.08 &  20.78$\pm$0.05 &  20.37$\pm$0.03 &  19.92$\pm$0.09 &  20.44$\pm$0.07  &  $0.35^{+0.05}_{-0.03}$ \\
  12 & MMJ\,045426.76 &      $>25.1$   &  27.32$\pm$0.11 &  27.71$\pm$0.57 &  25.53$\pm$0.08 &  23.83$\pm$0.05 &  26.80$\pm$4.24 &  22.39$\pm$0.34 &  21.95$\pm$0.13 &  21.77$\pm$0.10 &  21.37$\pm$0.36 &  21.25$\pm$0.15  &  $0.89^{+0.10}_{-0.10}$ \\
  15 & MMJ\,045328.86 &      $>25.1$   &      $>26.6$    &     $>25.8$     &      $>25.1$    &      $>24.2$    &     $>24.2$     &     $>20.1$     &  21.08$\pm$0.06 &  20.83$\pm$0.04 &  20.48$\pm$0.15 &  21.48$\pm$0.19  &  $1.87^{+0.62}_{-0.85}$ \\
  \bf 17 &\bf  MMJ\,045431.35 &\bf  24.45$\pm$0.05 &\bf  24.37$\pm$0.03 & \bf 24.23$\pm$0.03 & \bf 23.72$\pm$0.03 & \bf 23.31$\pm$0.04 & \bf 25.94$\pm$1.40 & \bf 21.13$\pm$0.17 & \bf 22.21$\pm$0.17 & \bf 21.97$\pm$0.12 & \bf 22.37$\pm$1.21 & \bf 22.07$\pm$0.33  &  $\bf{0.60^{+0.11}_{-0.11}}$ \\
  18 & MMJ\,045411.57 & 24.83$\pm$0.05 &  24.65$\pm$0.03 &  24.23$\pm$0.03 &  23.53$\pm$0.03 &  22.82$\pm$0.03 &  22.68$\pm$0.11 &  22.27$\pm$0.37 &  22.37$\pm$0.20 &  21.78$\pm$0.10 &  21.41$\pm$0.37 &  21.80$\pm$0.25  &  $0.74^{+0.06}_{-0.10}$ \\
  26 & MMJ\,045349.69 & 26.41$\pm$0.14 &  26.25$\pm$0.06 &  27.72$\pm$0.38 &  26.04$\pm$0.10 &  24.58$\pm$0.09 &  24.35$\pm$0.40 &      $>20.1$    &       -         &         -       &         -       &         -        &  $1.33^{+0.03}_{-0.10}$ \\
  27 & MMJ\,045421.17 & 25.42$\pm$0.10 &  25.51$\pm$0.04 &  24.46$\pm$0.04 &  23.56$\pm$0.03 &  23.11$\pm$0.04 &  23.68$\pm$0.22 &  18.82$\pm$0.07 &  20.07$\pm$0.02 &  19.83$\pm$0.02 &  19.15$\pm$0.05 &  20.52$\pm$0.08  &  $1.80^{+0.25}_{-0.26}$ \\
  28 & MMJ\,045354.06 & 25.61$\pm$0.08 &  25.19$\pm$0.03 &  24.83$\pm$0.05 &  24.68$\pm$0.04 &  24.35$\pm$0.07 &  24.36$\pm$0.34 &  21.63$\pm$0.43 &       -         &         -       &         -       &         -        &  $1.89^{+0.35}_{-0.84}$ \\
  29 & MMJ\,045442.54 & 25.00$\pm$0.09 &  26.16$\pm$0.05 &  25.71$\pm$0.12 &  24.61$\pm$0.04 &  23.86$\pm$0.05 &     $>24.2$     &  20.65$\pm$0.13 &  21.59$\pm$0.10 &  21.42$\pm$0.07 &  20.49$\pm$0.16 &  21.31$\pm$0.16  &  $1.09^{+0.15}_{-0.15}$ \\
\hline
\end{tabular}

\label{tab:optprops}
\end{minipage}
\end{table}
\end{landscape}

\subsubsection{Photometric redshifts of SMGs}  
\label{sec:borys}

\begin{figure}
\includegraphics[width=8.5cm]{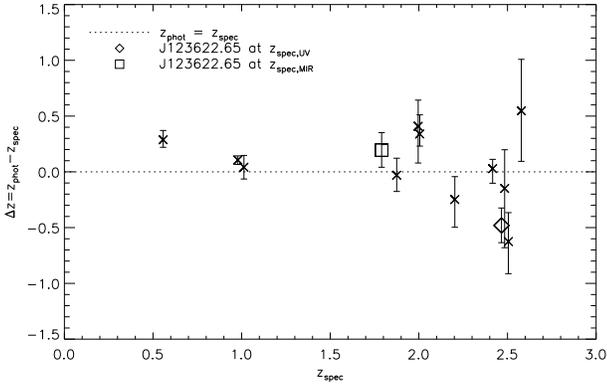}
\caption{Spectroscopic versus photometric redshifts for the SMGs in \citet{Borys05}. 
Galaxy SMMJ\,123622.65 has rest-frame UV/optical spectra with $z=2.466$ \citep{Chapman05, Swinbank04} but mid-IR 
analysis places it at $z=1.79\pm0.04$ \citep{Pope08}, therefore, SMMJ\,123622.65 is plotted at each redshift. 
Utilising 12-band photometry results in constraints that are sufficient for separating high-redshift SMGs from those at $z\sim 0.5$. 
}
\label{fig:deltz}
\end{figure}

Having located the \ms\ SMGs we now need to test whether any are potential cluster members. 
Photometric redshifts have been calculated for several sets of SMGs using various codes and 
spectral templates, generally designed for use on `normal' low redshift galaxies 
\citep[e.g.][]{Clements08}. However, SMGs
are located at high redshifts and powered by dusty starbursts which can be contaminated by AGN 
meaning that the stellar templates derived from low redshift galaxies may not be appropriate.   

Typically studies of SMGs do not have both spectroscopic and photometric information so testing 
of photometric redshift results is difficult. Often results from codes such as {\sc Hyperz} 
\citep{Bolzonella00} or {\sc Impz} \citep{Babbedge04} 
are compared to those from other, apparently cruder, methods of redshift estimation -- such as 
the radio to far-IR -(sub)-mm spectral
indices \citep[e.g.][]{Carilli99, Yun02, Aretxaga03, Clements08}. 
There are two main problems with this approach -- firstly, 
the comparison is made with a set of results which themselves are not known
to be correct, and secondly, the errors on the comparison redshifts are typically large, 
such that a general agreement can be confirmed but nothing more. In cases where 
spectroscopic information has been obtained for a sub-sample of galaxies typically they are 
small \citep[e.g.][]{Pope05}.

In this work we expand on previous analyses to examine how well redshifts can 
be constrained for a spectroscopically confirmed sample of typical SMGs (median $z\sim2.2$). 
We employ the sample of 12 SMGs from \citet{Borys05} 
which have spectroscopic redshifts from \citet{Chapman03, Chapman05} and \citet{Pope08}.
Photometry in 12 bands from {\it U} to 8\,$\umu$m is presented in  \citet{Borys05}.
We note that SMM\,J123622.65 has a rest-frame UV/optical spectrum with $z=2.466$ 
\citep{Chapman05, Swinbank04} but mid-IR spectroscopic analysis suggests $z=1.79\pm0.04$ \citep{Pope08},
therefore, we exclude it from the statistical analyses.
We also exclude SMM\,J123712.05 from the sample as it is undetected at 
wavelengths shorter than 2 $\umu$m and the power-law shape at longer wavelengths 
means that the redshift cannot be derived.

We use the {\sc Hyperz}\footnote{We use {\sc Hyperz} version 10.0 
(http://www.ast.obs-mip.fr/users/roser/hyperz/).} package \citep{Bolzonella00} for our photometric redshift estimates. 
The program calculates expected magnitudes in the observed filters for given 
SEDs of a library of model star-formation histories, 
with different ages, reddening and redshifts; the observed and expected magnitudes are compared in each 
filter. 
We use the \citet{Bruzual93} spectral templates provided with the 
{\sc Hyperz} package which represent star-formation histories resulting in SEDs which match
local ellipticals (E), Sb, a single burst (Burst) and a constant Star-formation rate (Im).
Redshifts between 0 and 7 are considered.
SMGs are known to be dusty systems, therefore, we initially allow reddening of 
$0\le A_V\le 5$, in steps of 0.2 using the \citet{Calzetti00} reddening law. 
Ages of the galaxies are required to be less than the age of the Universe at 
the appropriate redshift.

In Fig. \ref{fig:deltz}, we compare the spectroscopic with our photometric redshifts 
for the \citet{Borys05} galaxies; 
error bars shown are the {\sc Hyperz} quoted 99\% confidence intervals which 
represent the $1\sigma$ limits more realistically than the {\sc Hyperz} quoted 68\% intervals
for this sample. 
The average scatter in $\Delta z = z_{\rm phot} - z_{\rm spec}$ is $0.27\pm0.21$, 
although this is smaller for the $z\le 1.5$ SMGs ($\Delta z = 0.15\pm0.13$) than for the 
more distant sample ($\Delta z =0.32\pm0.22$),
demonstrating that when testing photometric redshifts of these galaxies it is 
important to consider a comparison sample with the same redshift distribution as the population under study.
Our results are unchanged if we restrict $A_V$ to less than 2.5 although a tighter limit than this
does affect the results -- increasing $\Delta z$. 

AGN contamination in the 8\,$\umu$m IRAC filter is possible for SMGs (Hainline, L. J. et al. in prep.), 
therefore we repeat the photometric redshift calculations as above but 
excluding the 8\,$\umu$m data from the analysis. For all the SMGs $\Delta z_{{\rm no 8 \umu m}}\le\Delta z_{{\rm all bands}}$ yielding an average scatter  
$\Delta z = 0.23\pm0.22$ for the high redshift ($z>1.5$) galaxies -- consistent with the view that the 8\,$\umu$m fluxes are contaminated.

We also find no change if we allow a full range of templates: 
E, S0, Sa, Sb, Sc, Sd, Burst, and Im spectral types, and  
minimal change if only we restrict the templates to only Burst 
and Im models -- despite the fact that when the full range of templates 
is allowed most of the galaxies are 
best-fit by the burst and elliptical models. We therefore find it 
unlikely that the addition of any further pure-stellar templates 
will improve the redshift accuracy.

In this work we are interested in separating low redshift ($z\sim0.5$; potential cluster member) SMGs from the typical high redshift ($z\sim2$) population, and it 
is clear from Fig. \ref{fig:deltz} that this is possible based on the photometric redshift resolution.

\subsubsection{Photometric redshifts of the AzTEC galaxy sample}
\label{sec:aztecphotoz}

\begin{figure*}
\includegraphics[width=17cm]{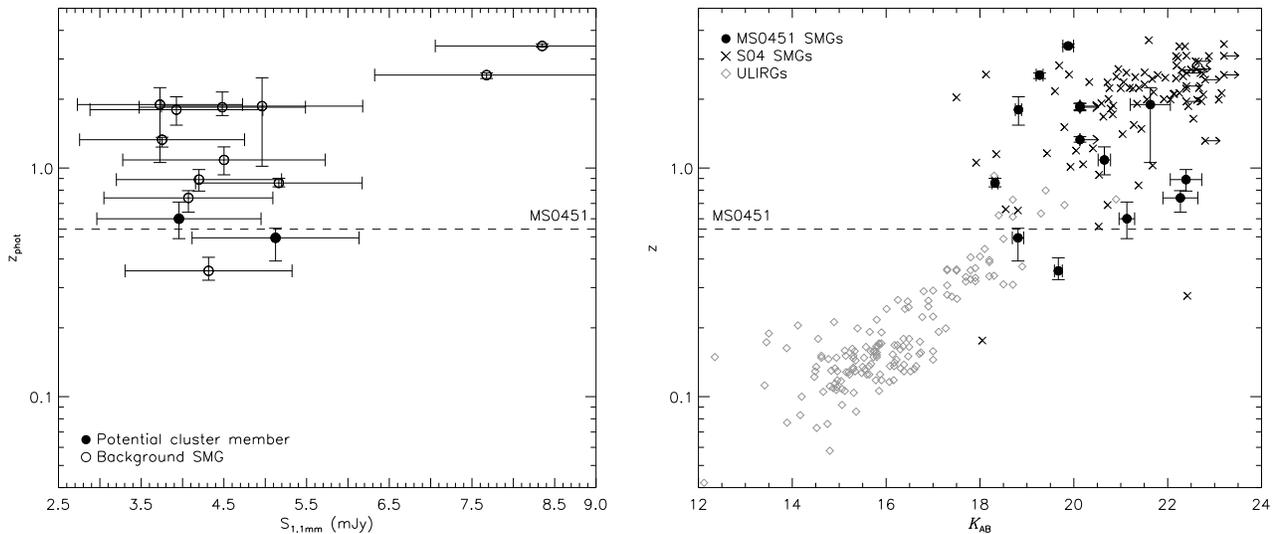}
\caption{
The left-hand panel shows photometric redshifts of SMGs against the 1.1 mm AzTEC fluxes.
Two SMGs -  MMJ\,045421.55 and MMJ\,045431.35 - have photometric redshifts consistent
with being cluster members and are highlighted;
we discuss these two sources in detail in \S\ref{sec:aztecphotoz}.
In the right-hand panel we show redshift versus $K$-band magnitudes of \ms\ SMGs compared to 
the spectroscopic sample presented in \citet{Smail04}, and local ULIRGs from \citet{Kim02} 
and \citet{Stanford00}. Both datasets occupy the same parameter space suggesting 
it is unlikely that any of our photometric redshifts are extreme outliers.
}
\label{fig:zvskmag}
\label{fig:zvsfluz}
\end{figure*}

We use {\sc Hyperz} to calculate the redshifts of the SMGs in the field of \ms\ from our optical, 
near-IR and mid-IR photometry. The \citet{Bruzual93} spectral templates included with
{\sc Hyperz} are used, and, as discussed in \S\ref{sec:borys}, we consider only the Burst, 
Elliptical, Sb and constant star-formation rate models, limit reddening to
$0\le A_V\le 5$, and redshifts to $0\le z\le 7$.
Bright, resolved galaxies with uncharacteristically high primary photometric redshifts 
are considered to lie at the calculated secondary solutions.

The redshift estimate of each counterpart is presented in Table \ref{tab:optprops}, and 
in Fig. \ref{fig:zvsfluz} we show the SMG photometric redshifts versus 
AzTEC 1.1-mm fluxes. Although, most of the AzTEC sources are high redshift 
background galaxies, two -- MMJ\,045421.55 and MMJ\,045431.35 -- are possible cluster members,
and are discussed further below. 
We note that as reported by other authors \citep[e.g.][]{Pope05} there appears to be a weak trend 
of redshift with millimetre flux -- the brightest millimetre galaxies
lie at higher redshifts.
Our sample has a median photometric redshift of 1.2 and an interquartile range of $z=0.7$--$1.9$ -- lower than spectroscopic 
studies \citep[e.g.][$\langle z\rangle=2.2$]{Chapman05} but similar to other photometric studies
\citep[e.g. $\langle z\rangle=1.4$;][]{Clements08}. We note that if we exclude the potential cluster 
members MMJ\,045421.55 and MMJ\,045431.35 from this analysis the median redshift of the field 
SMGs in this study is 1.8 and the interquartile range is $z=0.9$--$1.9$.

Although in \S\ref{sec:borys} we showed that the exclusion of 8\,$\umu$m
information can sometimes improve the accuracy of photometric redshift estimates of SMGs,
the redshifts reported in Table \ref{tab:optprops} include the 8\,$\umu$m photometry.
This is because for the galaxies with only IRAC detections the resulting lack 
of information about the location of the $1.6\umu$m stellar peak means that 
only weak redshift constraints are possible if we remove the 8 $\umu$m photometry.
Critically, the inclusion or exclusion of the 8\,$\umu$m information does not affect MMJ\,045421.55
and MMJ\,045431.35 which still both have photometric redshifts consistent with the cluster. 

In Fig. \ref{fig:bzk} we showed the $B-z$ versus $z-K$ colour-colour plot for AzTEC 
galaxies in \ms\ and used the {\it BzK} selection criteria \citep{Daddi04} to separate
galaxies above and below $z=1.4$. We similarly group the SMGs based on the full 
photometric analysis from {\sc Hyperz} and find that both methods broadly agree.
In Fig. \ref{fig:zvskmag} we show the redshifts and $K$-band magnitudes of \ms\ SMGs 
in comparison to the spectroscopic sample of SMGs examined by \citet{Smail04},
and local ULIRGs from \citet{Kim02} and \citet{Stanford00}.
The apparent {\it K}-band magnitudes of our AzTEC sample are consistent with 
their estimated redshifts, when compared with spectroscopic SMG surveys and 
local ULIRGs, suggesting that our photometric redshifts are reasonable.
The general agreement between our photometric redshift analysis, {\it BzK} and 
{\it K}-band magnitudes supports the derived redshifts of the AzTEC sources.

The potential cluster members, MMJ\,045421.55 and MMJ\,045431.35, are discussed in 
detail here. The other identified SMGs are examined in Appendix \ref{sourcenotes}.

\paragraph*{MMJ\,045421.55}

MMJ\,045421.55 is identified through radio emission 2.2 arcsec from the AzTEC centroid,
which lies between two optical galaxies (2.1 and 1.6 arcsec from the northern (C1) and southern (C2)
galaxies respectively). There is also an IRAC source at the location of the radio emission,
which appears slightly extended towards C2 (Fig. \ref{fig:cutouts}).
The three possible explanations for this system are:
C1 and C2 are interacting and both millimetre bright;  
C2 is the millimetre counterpart; or neither C1 nor C2 are responsible for the millimetre emission. 
Each of these possibilities is discussed below.

Individual photometric analysis of C1 and C2 (excluding the IRAC information) suggests that 
they are both potential cluster members, and with a separation of 2.6 arcsec - corresponding
to $\sim17$ kpc at $z=0.54$ -- they could be in the early stages of a merger.
Although high-resolution (0.1 arcsec) archival \HST\ 
F814W imaging shows no evidence of disturbance (Fig. \ref{fig:cutouts}), the 
expected tidal tails can be low surface brightness features, making them difficult to 
detect.
Such interactions are widely known to trigger dusty starbursts in which the radio emission
appears to be located between the optical nuclei \citep[e.g. systems
similar to VV 114;][]{Frayer99, LeFloch02, Iono04}.

With the aim of measuring redshifts we targeted C1 and C2 with the 
ISIS long-slit spectrograph on the WHT during service time in 2009 February.
The total integration time was one hour in $\sigma =1.1$-arcsec seeing 
and standard reduction techniques were employed. 
A faint continuum was observed (the two 
targets are blended), but no features suitable for redshift 
measurement were detectable. A possible faint ($\sim 2\sigma$) emission feature at 
7852 \AA\ is visible, which, if real, is most likely to be 
[O{\small III}] ${\rm \lambda 5007}$ \AA\ at $z=0.568$ -- placing the galaxy in a small 
group just behind the cluster. However, the feature is tentative and inconclusive.

On the assumption that C1 and C2 lie at the same redshift, and that the 
millimetre, radio and IRAC fluxes are emitted from a merging system
as described above, we combine the optical fluxes from C1 and C2.
The optical to near-IR photometric redshift of the whole system is $z=0.50^{+0.05}_{-0.10}$
which, as shown in Fig. \ref{fig:zvskmag}, agrees with the $z$ versus $K$
ULIRG trend. Therefore, if C1 and C2 are an interacting system,
MMJ\,045421.55 a possible cluster member.

The radio counterpart lies only $2.6\sigma$ from C2 (compared to $3.5\sigma$ from
C1), and the IRAC emission of MMJ\,045421.55 appears extended towards
C2. Therefore, in the situation where C1 and C2 are unassociated
we find it most likely that the C2 is the counterpart to the IRAC, radio and millimetre 
flux. In this case we obtain a photometric redshift of $z=0.51^{+0.07}_{-0.05}$ -- once again 
placing MMJ\,045421.55 in the region of \ms.

Based on the intrinsic faintness of many SMG counterparts it is feasible that the 
identified radio and IRAC emission arises from an optically faint 
background galaxy, unrelated to C1 and C2, and undetectable in our observations. 
Previous SMG surveys have confused low redshift galaxies with the source of
sub-mm emission due to the lensing of the background source \citep[e.g.][]{Chapman02}. 
In these situations it is only possible to distinguish
between a lensed background galaxy and a foreground cluster member through
the detection of the faint optical counterpart in deeper imaging, or by the detection of 
CO emission lines.

\paragraph*{MMJ\,045431.35}

The counterpart to MMJ\,045431.35 is securely identified 5.4 arcsec from 
the AzTEC centroid through its radio emission. The source is detected 
in all our \Spitzer\ and ground-based imaging and resolved
in the \HST\ F814W image into merging galaxies with centroids
separated by $\sim 1.6$ arcsec ($\sim 10$ kpc at $z=0.54$) and
tidal tails between them (Fig. \ref{fig:cutouts}). 
We suggest that a dusty starburst triggered by the interaction
between the two galaxies is causing the millimetre emission from this system. 
The photometric redshift is calculated as $z=0.60\pm0.11$, making MMJ\,045431.35
a possible cluster member. 
\\

If MMJ\,045421.55 and MMJ\,045431.35 are cluster members at $z=0.54$ SED-fitting
suggests they each have $L_{\rm FIR}\sim5\times10^{11}L\sun$ and thus SFR$\sim50$ \myr. 
We also find that they are colder than typical $z\sim2$ SMGs, with dust temperatures of
$T_d=15\pm4$ K (for $\beta=1.5$), or $T_d=30\pm5$ K (for $\beta=1.1$), compared to 
$T_d\sim40$ K and $\beta=1.5$ for archetypal SMGs.
Such properties are not unprecedented -- 
the spectroscopic survey of SMGs by \citet{Chapman05} contained examples of sources
with equivalent millimetre-to-radio flux ratios at $z\sim0.5$, suggestive of
galaxies containing cold dust. 
Similarly, in the SCUBA Local Universe Galaxy Survey (SLUGS)
\citet{Dunne00a} surveyed local {\it IRAS}-bright galaxies with SCUBA
at 850\,$\umu$m and utilised the combined {\it IRAS} and SCUBA photometry 
for SED fitting. This sample of local galaxies has $T_d=35.6\pm4.9$ K,
and $\beta=1.3\pm0.2$. 
Indeed, cold low-redshift galaxies are easier to detect at 850\,$\umu$m or 1.1-mm 
than their hotter counterparts. This is because colder dust produces emission
which peaks at longer wavelengths than hot dust. Therefore, we do not find it unreasonable
that one or both of MMJ\,045421.55 and MMJ\,045431.35 are cluster members
at $z=0.54$ with $T_d\sim30$ K and $\beta=1.1$.

If both MMJ\,045421.55 and MMJ\,045431.35 are members of \ms, their combined 
SFR is $\sim100$ \myr\ -- a significant fraction of the SFR of all the cluster
galaxies within 2 Mpc \citep[$200\pm100$\myr;][]{Geach06}
MMJ\,045431.35 lies $\sim 2.6$ Mpc from the cluster centre (about half of the
turnaround radius), but MMJ\,045421.55 is much closer to the centre: $\sim 1$ Mpc in projection. 
Notably, both of these systems are most likely mergers. Although not conclusive, 
if they are both cluster members, 
this suggests that existing starbursts are not instantaneously suppressed as
they are accreted into the cluster environment, and can survive until at least
1 Mpc from the cluster centre.
Alternatively the galaxy pairs could have been accreted into the cluster -- suggesting
that accreted galaxies can retain their gas reserves during infall into clusters \citep{Geach09}.

\subsection{Cluster and field SEDs}
\label{sec:seds}

\begin{figure*}
\begin{minipage}{17.5cm}
\includegraphics[width=17.5cm]{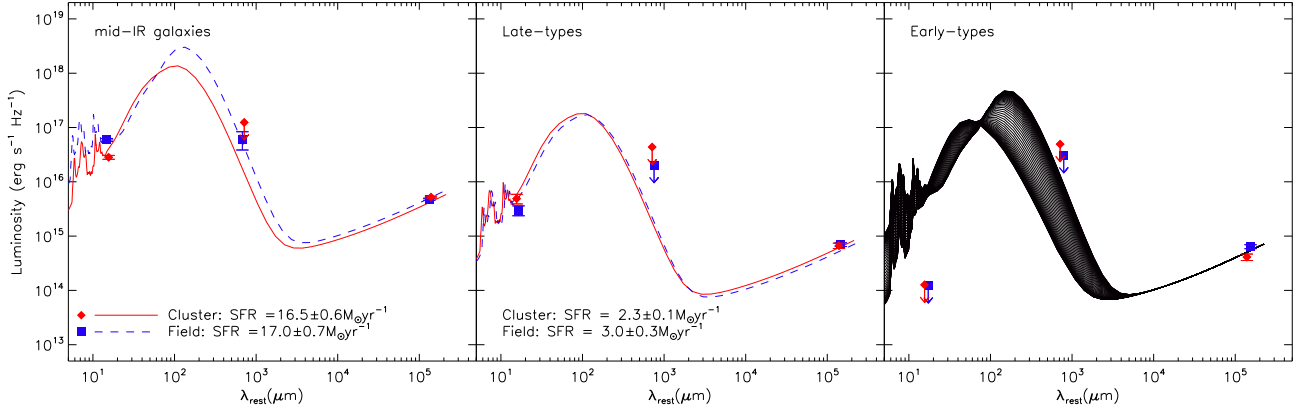}
\caption{Clipped weighted-average SEDs of the 24$\umu$m, late- and early-type populations of \ms\ and the field. 
In the case of non-detections the 3$\sigma$ noise limit is marked by arrows. 
The field populations are calibrated to their mean redshifts of 0.58, 0.46, and 0.39 for the mid-IR galaxies, late- and early-types and respectively.
We also show the best-fitting SEDs and corresponding SFRs \citep{Kennicutt98} from \citet{DaleHelou02} for the mid-IR and late-type galaxies; 
since the early-type populations are only detected at 1.4 GHz we cannot select one best-fitting SED and instead display all 
the \citet{DaleHelou02} SEDs on this panel.
Although the observed populations have similar activity levels, in the late-type galaxies this is due to the flux limits of our observations
and the different luminosity distances between the samples. If the cluster and field late-type populations were equivalent, the cluster
should have ${\rm SFR}\sim3$ times higher than observed. We deduce that the cluster environment has caused a reduction in the SFR of this population,
and that these galaxies are probably in the process of transformation onto the red sequence.
}
\label{fig:seds}
\end{minipage}
\end{figure*}

\begin{table*}  
\begin{minipage}{160mm} 
\caption{Parameters and results of stacking on known cluster and field galaxies in the 24\,$\umu$m, 1.1 mm and 1.4 GHz images.
SFRs are based on SED fits, except for the early-type galaxies, which are based purely on 24 $\umu$m detection limits. 
Non-detections are represented by $3\sigma$ limits. 
Since the samples are flux limited and have various median redshifts we also present the expected observed SFR of the field  
mid-IR and late-type samples were they to have $z_{median}=0.54$ (\S\ref{sec:seds}). Differences between this value and the observed cluster 
SFR for the late-type galaxies suggests that the cluster late-type population is less active than the field late-type population.}
\setlength{\tabcolsep}{1mm}
\begin{tabular}{lccccccccc}
\hline
Population & ${\rm N_{24\umu m}}$\footnote{The number of galaxies used in the stacking -- excluding those close to the edge, off-image, or near detected sources} & ${\rm S_{24\umu m}}$\footnote{Clipped weighted-average flux} & ${\rm N_{1.1mm}}^a$ & ${\rm S_{1.1mm}}^b$ & ${\rm N_{1.4GHz}}^a$ & ${\rm S_{1.4GHz}}^b$ & $z_{\rm median}$ & SFR &  SFR ($z=0.54$) \\
 & & ($\umu$Jy) &  & ($\umu$Jy) &  & ($\umu$Jy) & & (\myr) & (\myr) \\
\hline
Cluster mid-IR galaxies& 14  & $222\pm 18$ & 14  & $<970$ & 15  & $41.1\pm 1.6$ & 0.54 & $16.5\pm0.6$  & $16.5\pm0.6$ \\
Field mid-IR galaxies  & 84  & $331\pm 21$ & 94  & $340\pm 130$ & 89  & $26.8\pm 1.1$ & 0.58 & $17.0\pm0.7$ & $12.6\pm2.8$ \\
\hline
Cluster late-type      & 167 & $38.4\pm 7.7$ & 131 & $<340$ & 162 & $5.2\pm 0.6$ & 0.54 & $2.3\pm0.1$ & $2.3\pm0.1$ \\
Field late-type        & 315 & $34.4\pm 7.1$ & 309 & $<230$ & 355 & $8.0\pm 0.6$ & 0.46 & $3.0\pm0.3$ & $7.8\pm1.1$ \\
\hline
Cluster early-type     & 148 & $<1.0$ & 106  & $<390$ & 144 & $3.2\pm 0.4$ & 0.54 & $<0.07$ & $<0.07$ \\
Field early-type       & 35  & $<2.1$ & 32  & $<530$ & 40  & $10.9\pm 0.9$ & 0.28 & $<0.03$ & $<0.19$ \\
\hline
\end{tabular}
\label{tab:stackparam}
\end{minipage}
\end{table*}

We can also investigate obscured star-formation of the general galaxy population in \ms\ below 
the flux limit of our AzTEC map by stacking fluxes at the positions 
of known cluster galaxies.
Optical galaxies in \ms\ were morphologically classified by \citet{Moran07b} using the 
scheme defined by \citet{Abraham96}, which we group into early-types 
and late-types for this analysis. 
We also define a mid-IR sample of galaxies which are bright at 24\,$\umu$m, based on 
the catalogue of \citet{Geach06} with $S_{24\mu m}\ge200\ \umu$Jy. 
To reduce contamination we consider only those galaxies with 
spectroscopic redshifts and to search for environmental dependencies we also examine spectroscopically identified field galaxies. 
The cluster members are required to have $0.52<z<0.56$ and the field population is outside this window.
The field samples have median redshifts of 0.58, 0.46, and 0.28  
with interquartile ranges of $z=$ 0.33--0.82, 0.26--0.62, and 0.20--0.51
for the mid-IR galaxies, late-types and early-types respectively.
For each of these six samples we stack the MIPS 24\,$\umu$m, AzTEC 1.1 mm and VLA 1.4GHz fluxes. 
We note that the both galaxy samples are optical magnitude limited due to the requirement for a spectroscopic redshift. 
Therefore, the stacked SEDs may not be representative of the entire population, in particular the most obscured galaxies
are likely to fall below the optical magnitude limit. However, we expect such selection effects to equally affect the 
cluster and field samples allowing us to compare populations between the two density regimes.

Any galaxies within 9 arcsec (the radius of the AzTEC beam) of an AzTEC map pixel with S/N$\ge\pm3.5$ 
are removed prior to stacking the AzTEC map. The contribution 
from each AzTEC-faint galaxy is weighted by the inverse of the squared noise at that pixel to 
calculate the weighted mean 1.1 mm flux of each population.
Similarly we calculate the clipped weighted mean radio flux of each sample using our 
1.4 GHz VLA map, correcting for bandwidth smearing, and stack 24\,$\umu$m image in the same way.
The stacking parameters are given in Table \ref{tab:stackparam} and the resulting SEDs are presented 
in Fig. \ref{fig:seds}.

We fit the mid-IR and late-type galaxies with template SEDs from \citet{DaleHelou02} 
and calculate the corresponding SFRs based on the far-IR luminosity 
\citep{Kennicutt98}. Mid-IR cluster galaxies without bright AzTEC counterparts have ${\rm SFR=16.5\pm0.6}$ \myr\ 
which is consistent the SFR estimate of $17.0\pm0.7$ \myr for the mid-IR field population. 
In contrast, the late-type galaxies in the field have ${\rm SFR=3.0\pm0.3}$ \myr, 
compared to ${\rm SFR=2.3\pm0.1}$ \myr\ in the cluster. 
Both cluster and field early-type populations are undetected at 1.1 mm and 24\,$\umu$m 
and SED fitting to the 24\,$\umu$m limits suggest, on average, SFR$<0.06$ \myr.

Since our samples are flux limited, the different redshift distributions will
affect the derived SFR.
The extent of this effect on our results is tested by calculating an observed 
SFR for each sample by fitting templates with $S_{24\umu m}=200$ $\umu$Jy 
at the redshift of the galaxies included in the stacks.
Due to the generally higher redshift of the field sample, we find that identical 
mid-IR populations would be observed with SFR $0.74\pm0.16$ times lower in the cluster 
than the field. In fact the stacked cluster population is observed with SFR $0.97\pm0.05$ times lower
than the field population, therefore, to within $\sim1\sigma$ the mid-IR samples
are consistent. 
However, the equivalent test for late-type galaxies, where the field sample is, 
on average, lower redshift, suggests that, due to the different luminosity distance,
identical populations in our analysis would appear 2.6 times more active in the cluster
than in the field. 
In fact, the  cluster late-type galaxies are less active than the field.
Therefore, the cluster environment is suppressing star-formation activity 
in the late-type population.

\citet{Geach06} estimated the total SFR within 2 Mpc of the core of \ms\ as $200\pm100$ \myr, 
based on converting 24\,$\umu$m fluxes of colour-selected 24-$\umu$m detected galaxies 
to total IR luminosities with the 
{\it average} SED from \citet{DaleHelou02}. Our analysis presented in this paper includes radio and 1.1 mm data enabling 
us to better characterise the cluster populations, and by stacking we can probe a fainter population.
Within 2 Mpc of cluster core we calculate SFR$>315\pm50$~\myr\ from the 
spectroscopically confirmed mid-IR and late-type galaxies. 
To this we can then add $50$~\myr\ for MMJ\,045421.55, the potential ULIRG within 
2 Mpc of the cluster core.

\section{Conclusions}
\label{sec:conc}

In this study we have investigated the dust-obscured star-forming population 
of the galaxy cluster \ms; we utilise 1.1-mm observations to 
study obscured star-formation in \ms. We present a $\sigma\sim 1.1$ mJy AzTEC map 
of the central 0.1 deg$^2$ of \ms, within which 36 sources are 
detected at S/N$\ge3.5$. We use radio, 24\,$\umu$m, IRAC and SMA observations to 
precisely locate 18 of these SMGs.

We calculate the reliability of photometric redshifts for SMGs and find that they are 
able to remove the bulk of background contamination, allowing us to isolate potential 
cluster members using our optical, near- and mid-IR photometry.
Based on these redshifts we find two SMGs which are possible cluster members: MMJ\,045421.55
and MMJ\,045431.35.  
These systems are both resolved into close pairs of galaxies by our ground-based and \HST\ imaging,
suggesting that interactions have triggered their starbursts.
If they are cluster members both of these SMGs contain cold
dust with $T_d\sim30$ K, for $\beta=1.1$ \citep[similar to SLUGS galaxies;][]{Dunne00a}
and have SFRs of $\sim50$ \myr\ each. 

\citet{Geach06} compared obscured activity, based on $24 \umu$m emission, in \ms\ and
\cl\ at $z=0.39$, and found that \cl\ has ${\rm SFR}\sim5$ times that of \ms\ within 
2 Mpc of the clusters centres. They show that, taking into account the slightly different
redshifts and masses of these two structures, \ms\ is underactive and \cl\ overactive
at $24 \umu$m. Therefore, it is likely the other galaxy clusters, including \cl,  
could contain ULIRGs in significantly larger numbers than we find in \ms.

To further investigate the obscured star-forming population which lies below the limit of our AzTEC 
observations we create composite SEDs of spectroscopically confirmed mid-IR, early- and late-type
cluster members and compare them to the corresponding field populations. 
As expected we find that both early-type populations are undetected at both 24\,$\umu$m and
1.1 mm and so are unlikely to be actively forming large numbers of stars ($<0.1$ \myr).
The  24\,$\umu$m galaxies are significantly more active than 
the morphologically classified late-types with SFRs $\sim15$ \myr\ versus $\sim3$ \myr\ on average. 
We find that the star-formation activity in the cluster late-type population, compared to a
redshift-matched field population is quenched and $\sim3$ times lower than expected.
Mid-IR galaxies do not show this trend suggesting the more intense activity in these systems is 
more robust to environmental influences.
We find that the total ${\rm SFR>315\pm50}$ \myr\ in the central 2 Mpc of \ms. However, if MMJ\,045421.55
is a cluster member it has ${\rm SFR}\sim50$ \myr\ and lies 1 Mpc from the cluster centre, taking
the total SFR within 2 Mpc to $\ga360$ \myr.

\section*{Acknowledgements}

We thank an anonymous referee for helpful comments which greatly improved the clarity of this paper.
We would also like to thank C. J. Ma and Harald Ebeling for providing us with reduced CFHT {\it U}-band images, 
Richard Ellis for supplying reduced \HST\ WFPC2 images, and Sean Moran for the usage of spectroscopic 
catalogues and attributed information.

J.~L.~W., I.~R.~S and J.~E.~G. acknowledge support from the Science and Technology Facilities Council (STFC),
and K.~E.~ K.~ C. acknowledges support from an STFC Fellowship. 
K.~S was supported in part through the NASA GSFC Cooperative Agreement NNG04G155A.
Support for this work was provided in part by the NSF grant AST 05-40852 and the grant from the 
Korea Science \& Engineering Foundation (KOSEF) under a cooperative agreement with the Astrophysical 
Research Center of the Structure and Evolution of the Cosmos (ARCSEC).
Additional support for this work was provided by NASA through an award issued by JPL/Caltech.

The James Clerk Maxwell Telescope is operated by The Joint Astronomy Centre on behalf of the Science and Technology Facilities Council of the United Kingdom, the Netherlands Organisation for Scientific Research, and the National Research Council of Canada.
This work is based in part on observations made with the \textit{Spitzer Space Telescope}, which is operated by the Jet Propulsion Laboratory, California Institute of Technology under a contract with NASA. This work also made use of the Spitzer Archive, which is operated by the Spitzer Science Center.
Based in part on data collected at Subaru Telescope, which is operated by the National Astronomical Observatory of Japan.
Based on observations obtained with MegaPrime/MegaCam, a joint project of CFHT and CEA/DAPNIA, at the Canada-France-Hawaii Telescope (CFHT) which is operated by the National Research Council (NRC) of Canada, the Institute National des Sciences de l'Univers of the Centre National de la Recherche Scientifique of France, and the University of Hawaii.
The William Herschel Telescope and its service programme are operated on the island of La Palma by the Isaac Newton Group in the Spanish Observatorio del Roque de los Muchachos of the Instituto de Astrof{\'i}sica de Canarias.

\appendix
\section{Notes on Individual Sources}
\label{sourcenotes}

\begin{figure*}
\begin{minipage}{17.5cm}
\includegraphics[width=18cm]{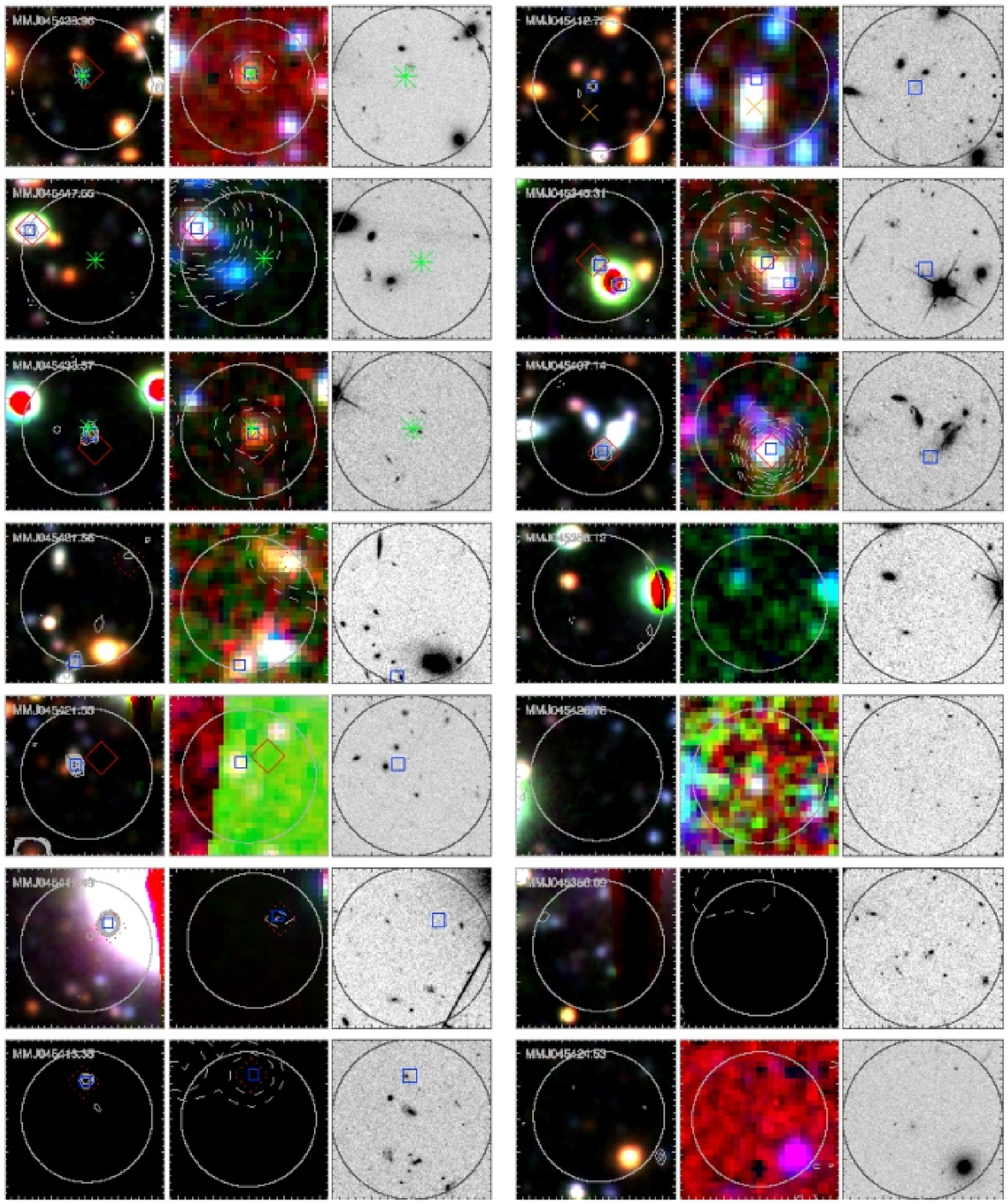}
\caption{$25\times25$ arcsec images centred on each AzTEC galaxy are shown in the 
left-hand and middle panels; north is up and east is to the left. 
In the left-hand panel we show Subaru data: {\it BVR} colour images with solid 
radio contours at -3 (dotted), 3, 4, 5, 6, 7, 8, 9, 10$\times\sigma$ are overlayed.
The middle panel contains true-colour images containing the IRAC 
$3.6+4.5$ (blue), 5.8 (green) and 8.0\,$\umu$m (red) data; dashed contours present 
24\,$\umu$m flux at 5, 10, 20, 30, 40, 50, 100, 200, ...1000$\times\umu$Jy 
per pixel and the right-hand panel contains $20\times20$ arcsec \HST\ F814W cutouts centered 
at the AzTEC position.
In all images circles are centred on the AzTEC positions and have 
20 arcsec diameter, corresponding to our radio and mid-IR search radius. 
24$\umu$m and radio counterparts are highlighted with diamonds and squares respectively; 
dotted symbols represent tentative counterparts (with $0.05\le P\le 0.10$) and 
the sizes of the symbols are representative of the typical astrometric errors of these data.  
We mark the positions of SMA detected sources with stars and X-ray sources \citep{Molnar02} 
with crosses.
}
\label{fig:cutouts}
\end{minipage}
\end{figure*}

\begin{figure*}
\begin{minipage}{17.5cm}
\includegraphics[width=17.5cm]{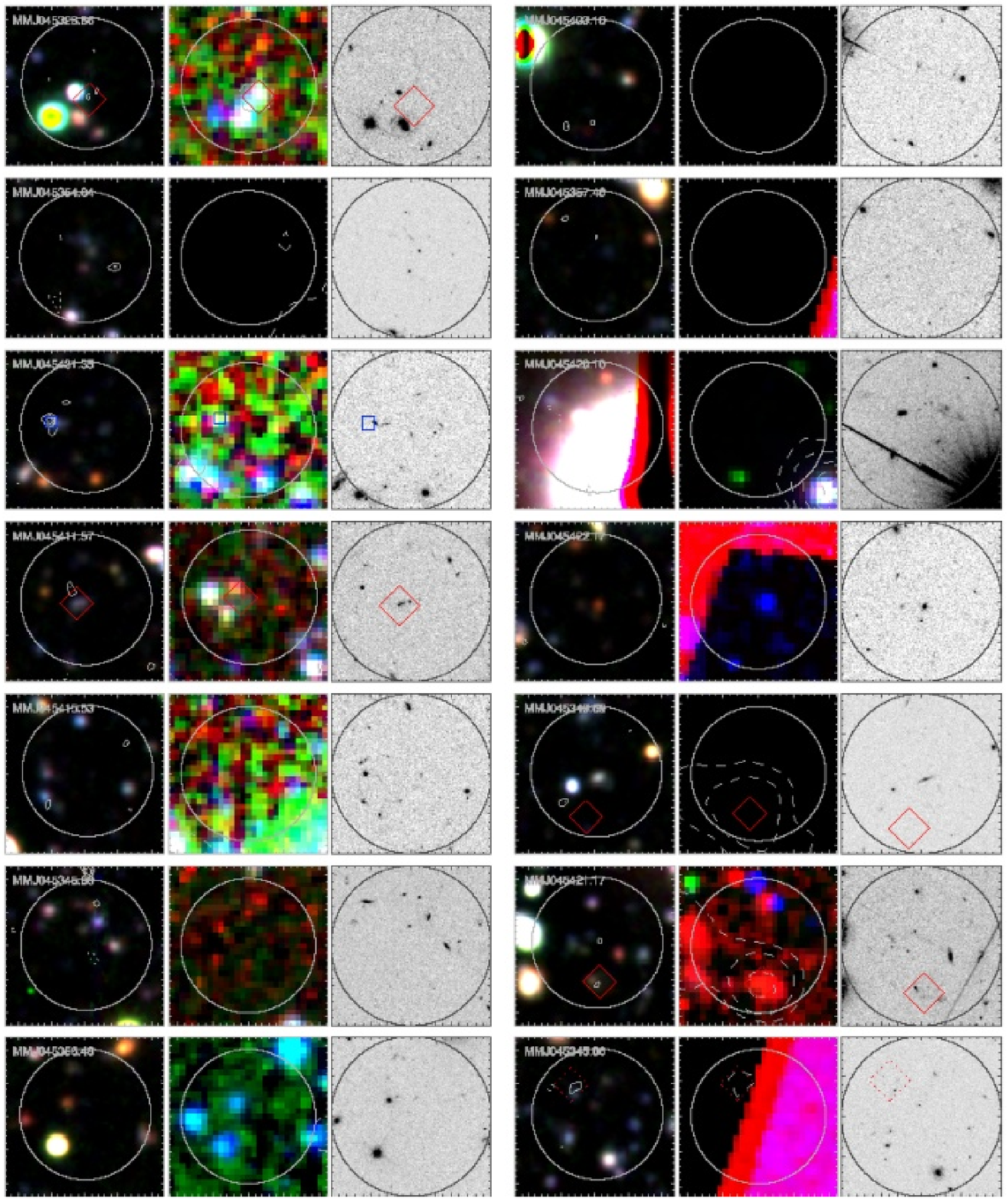}
\contcaption{}
\end{minipage}
\end{figure*}

\begin{figure*}
\begin{minipage}{17.5cm}
\includegraphics[width=17.5cm]{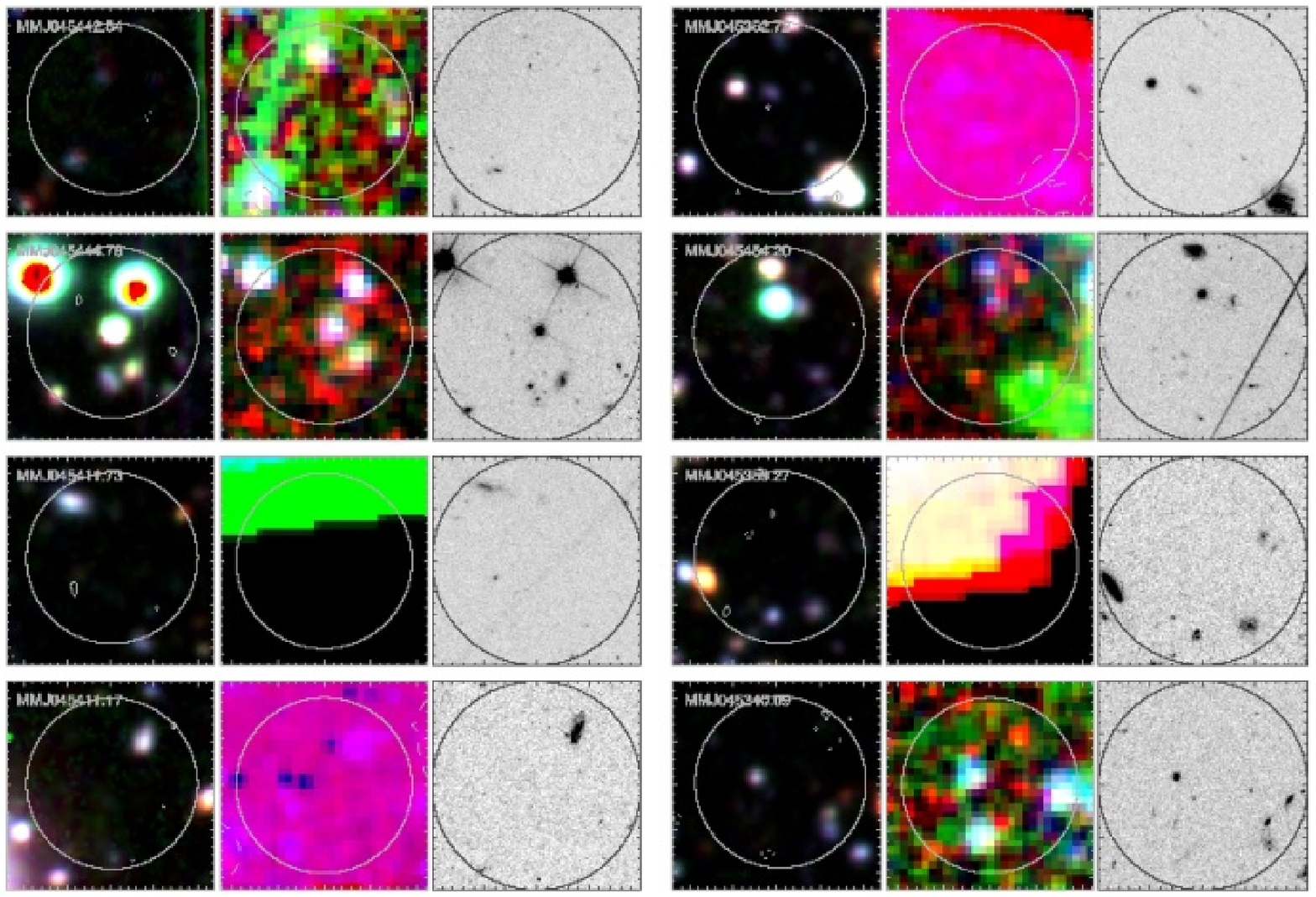}
\contcaption{}
\end{minipage}
\end{figure*}

In Fig. \ref{fig:cutouts} we present images centered on each 
AzTEC source, in order of decreasing S/N. We show Subaru and IRAC colour, and \HST\
images, with radio and 24 $\umu$m contours. Identified galaxies are labelled and discussed below.\\
{\bf 1. MMJ\,045438.96:} SMA observations of this galaxy confirm the identified radio and mid-IR 
counterparts. The corresponding red optical and IRAC source at 
$z_{\rm phot}=3.42^{+0.07}_{-0.05}$, has a disturbed morphology in the \HST\ image.\\ 
{\bf 2. MMJ\,045447.55:} This galaxy has been located with the SMA but does not have any 
radio or 24\,$\umu$m counterparts, or optical or IRAC galaxies at the SMA 
position. Therefore, MMJ\,0455447.57 could be one of a population of very distant
SMGs, or be much cooler and more obscured than typical SMGs.\\ 
{\bf 3. MMJ\,045433.57:} SMA observations confirm the identified radio 
and mid-IR counterparts, corresponding to an optically faint but near-IR bright galaxy with 
$z_{\rm phot}=2.55^{+0.06}_{-0.09}$. The {\it HST} image shows three components 
to this galaxy such that a merger is the likely cause of the ULIRG. \\ 
{\bf 4. MMJ\,045431.56:} 
We identify a robust extended radio counterpart 9.5 arcsec from the 1.1 mm centroid, 
which also corresponds to faint mid-IR emission; we consider the corresponding red galaxy 
at $z_{\rm phot}=0.86^{+0.04}_{-0.03}$ to be the likely optical counterpart. 
The {\it HST} image of the region shows a galaxy with a bright core and a 
low surface-brightness tail or edge on spiral arm.
However, the picture of this SMG is complicated by the presence of a statistically 
tentative mid-IR counterpart 9.1 arcsec from the 1.1 mm centroid which is unassociated with the 
aforementioned radio emission. The 24\,$\umu$m position is coincident with an extended red
IRAC galaxy but no optical sources.
Further complexity arises from the SMA observation; MMJ\,045431.56 was targeted
but not formally detected. However, the SMA data does show two potential sources, 
close to an extended $\sim3\sigma$ peak in the radio map, $\sim3.7$ arcsec from the AzTEC centroid 
and unassociated with both the radio and 24\,$\umu$m counterparts previously discussed.
There is also a red IRAC source at this position, and the \HST\ image shows two optical galaxies.
However, in the IRAC and ground-based optical data the counterpart is blended with a
nearby elliptical galaxy.\\
{\bf 5. MMJ\,045421.55:} We find reliable coincident radio and 24\,$\umu$m counterparts, 
corresponding to two optical galaxies separated by 2.6 arcsec, which are possible 
interacting cluster members ($z_{phot}=0.50^{+0.05}_{-0.10}$). 
The system is discussed in detail in \S\ref{sec:aztecphotoz}.\\  
{\bf 6. MMJ\,045417.49:} Both mid-IR and strong radio counterparts are detected although the presence 
of a nearby, bright star prevents multi-wavelength optical and near-IR
study. Although the stellar halo is less extended in the {\it HST} image no counterpart is visible.\\ 
{\bf 7. MMJ\,045413.35:} 
The radio and 24\,$\umu$m identification agrees with a 2--3$\sigma$ peak in the 890\,$\umu$m `dirty' SMA 
map. Unfortunately this source lies outside of the IRAC and much of the 
ground-based optical coverage, although a galaxy is detected in the {\it U}-band and \HST\ imaging.\\ 
{\bf 8. MMJ\,045412.72:} As potentially part of the gravitationally lensed arc from \citet{Borys04b} this source is discussed fully in Wardlow et al. (2009b). 
Our optical imaging is too shallow to detect the ERO discussed in \citet{Borys04b}, but radio and extended IRAC emission betrays the likely counterpart. \\
{\bf 9. MMJ\,045345.31:} There are two secure radio and one 24\,$\umu$m counterparts. 
One of the radio galaxies corresponds to the 24\,$\umu$m position. 
The IRAC source has a red component at the mid-IR/radio position, but it blended with the emission from an optically 
saturated star making further conclusions difficult.
It is unclear whether the 24\,$\umu$m and radio flux is from the star or a background SMG counterpart which may be the cause of the red 
component of the IRAC emission.\\
{\bf 10. MMJ\,045407.14:}  
We find coincident reliable radio and 24\,$\umu$m counterparts which are located in a
red region at the edge of a bright, resolved galaxy which appears disturbed in the \HST\ image
and has $z_{\rm phot}=0.35^{+0.05}_{-0.03}$.
The morphology of the galaxy suggests that a merger has triggered a dusty starburst region, causing the millimetre emission.
The large size and optical brightness support the low redshift of this galaxy, although if the photometric
redshift errors are underestimated it could potentially be a cluster member.
Similarly, it is also possible that a background galaxy is the source of the millimetre emission.
The detection of CO emission lines is required to confirm either scenario.
\\ 
{\bf 12. MMJ\,045426.76:} There are no radio or mid-IR counterparts identified. However, an IRAC galaxy has colours suggestive of SMG emission
\citep{Yun08}, yielding a counterpart with $z_{\rm phot}=0.89\pm0.10$.\\
{\bf 15. MMJ\,045328.86:} A 24\,$\umu$m counterpart is detected 3.8 arcsec from the 
AzTEC position, corresponding to an IRAC source, but no detectable optical flux.
The IRAC photometry yields  $z_{\rm phot}=1.87^{+0.62}_{-0.85}$.
There is coincident faint radio emission ($\sim3\sigma$) at the position of the 24\,$\umu$m 
counterpart, increasing the likelihood that this is the correct identification. \\  
{\bf 17. MMJ\,045431.35:} A radio counterpart 4.8 arcsec from the AzTEC position coincides with an pair of merging galaxies which are resolved in the {\it HST} image and have
$z_{\rm phot}=0.60\pm0.11$; this possible cluster ULIRG is discussed further in \S\ref{sec:aztecphotoz}.\\  
{\bf 18. MMJ\,045411.57:} 24-$\umu$m emission betrays the source of the millimetre emission as a 
galaxy with $z_{\rm phot}=0.74^{+0.06}_{-0.11}$, which is 
resolved in the {\it HST} image as an interacting pair, separated by $<1$ arcsec.
There is faint radio emission ($\sim3\sigma$) $\sim 2.3$ arcsec to the northeast of the 24\,$\umu$m identification, 
coincident with a red IRAC source. We conclude that the interaction between the optical galaxy pair 
is producing all of the emission. However, it is possible that the radio and IRAC source is unrelated to the
optical emission, in this case we can only constrain the redshift to $z_{\rm phot}>1.14$.\\
{\bf 26. MMJ\,045349.69:} A faint optical galaxy with $z_{\rm phot}=1.33^{+0.03}_{-0.10}$ corresponds to the robust mid-IR counterpart. 
The galaxy is undetected by {\it HST} and the region is not covered by our IRAC mosaic.\\ 
{\bf 27. MMJ\,045421.17:} The mid-IR counterpart is coincident with both faint radio emission
and a faint red galaxy at $z_{\rm phot} = 1.80^{+0.25}_{-0.26}$, 
which the {\it HST} imaging resolves into a bright nuclear region and some extended emission.\\ 
{\bf 28. MMJ\,045345.06:} The galaxy is tentatively identified through its mid-IR emission, which
is coincident with faint radio flux; our IRAC mosaic does not cover the counterpart, but optical
photometry yields $z_{\rm phot} = 1.89^{+0.35}_{-0.84}$.\\
{\bf 29. MMJ\,045442.54:} Although there are no radio or mid-IR counterparts MMJ\,045442.54 is identified from its IRAC colours and the photometry of the 
corresponding faint galaxy yields  $z_{\rm phot}=1.09\pm0.15$.\\ 

\label{lastpage}

\end{document}